\documentstyle[aas2pp4,psfig]{article}

\received{24 February 1998}
\accepted{11 May 1998}
\journalid{506}{10 October 1998}

\slugcomment{To appear in ApJ number 506}

\lefthead{Rosa A. Gonz\'alez et al.}
\righthead{The Opacity of Galaxies}

\def\arcsper{\ifmmode \rlap.{^{\prime\prime }}\else
$\rlap{.}{^{\prime\prime}} $\fi}
\def\arcmper{\ifmmode \rlap.{^{\prime }}\else $\rlap{.}{^\prime} $\fi}
\def\degper{\ifmmode \rlap.{^{\circ }}\else $\rlap{.}{^\circ} $\fi}

\def\deg{\ifmmode^\circ\else$^\circ$\fi}

\def\H2{H$_2$}

\def\13co{$^{13}$CO}
\def\nh3{NH$_3$}
\def\lax{${_<\atop^{\sim}}$}
\def\gax{${_>\atop^{\sim}}$}

\def\cm2{cm$^{-2}$}
\def\cm3{cm$^{-3}$}

\def\sref#1 #2 #3 #4 {#1, #2, #3, #4}

\def\plotone#1{\centering \leavevmode
    \epsfxsize=0.85\columnwidth \epsfbox{#1}}

\def\plotfiddle#1#2#3#4#5#6#7{\centering \leavevmode
    \vbox to#2{\rule{0pt}{#2}}
    \includegraphics{#1}}
 
\begin{document}

\title {THE OPACITY OF NEARBY GALAXIES \\
FROM COLORS AND COUNTS OF BACKGROUND GALAXIES:\\[0.2in]
I. The Synthetic Field Method and its Application \\
to NGC 4536 \& NGC 3664\altaffilmark{1}}

\author {Rosa A.\ Gonz\'alez, Ronald J.\ Allen, Boris Dirsch\altaffilmark{2},\\
Henry C.\ Ferguson, Daniela Calzetti, and Nino Panagia\altaffilmark{3}}
\affil {Space Telescope Science Institute, Baltimore, MD 21218\\
e-mail: ragl,rjallen,ferguson,calzetti, \& panagia @stsci.edu;\\
bdirsch@astro.uni-bonn.de}
\altaffiltext{1}
{Based on observations with the NASA/ESA Hubble
Space Telescope obtained at the Space Telescope Science Institute, which is
operated by the Association of Universities for Research in Astronomy, Incorporated, under NASA contract NAS5-26555.}
\altaffiltext{2}
{Sternwarte der Universit\"at Bonn, D-53121 Bonn, FRD}
\altaffiltext{3}
{On assignment from Astrophysics Division, Space Science Department of ESA}

\begin{abstract}

We describe a new, direct method for determining the opacity of
foreground galaxies which does not require any {\em a priori}
assumptions about the spatial distribution or the reddening law of the
obscuring material.  The method is to measure the colors and counts of
background galaxies which can be identified through the 
foreground system. The method is calibrated, and the effects of confusion and obscuration are decoupled
by adding various versions of a suitable
deep reference frame containing only field galaxies with known
properties into the image of the foreground galaxy, and analyzing these
``synthetic field'' images in the same way as the real images.

We test the method on HST WFPC2 archived images of two galaxies which
are quite different: NGC 4536 is a large Sc spiral, and NGC 3664 is a
small Magellanic irregular. The reference frames are taken from the Hubble
Deep Field.

From the background galaxy counts, NGC 4536 shows an extinction $A_I \sim$
1 mag in the northwestern arm region, and lower than 0.5 mag in the
corresponding interarm region (no correction for inclination has been
attempted).  However, from the galaxy colors, the same reddening of
$E(V - I) \sim 0.2$ is observed in both the arm and the interarm regions.
In the interarm region, the combination of 
extinction and reddening can be explained by  
a diffuse component with a Galactic reddening
law ($R_V \approx 3$). 
In the spiral arm, however, the same diffuse, low opacity component
seems to coexist with regions of much higher opacity.  
Since the exposures are shorter the
results for NGC 3664 are less clear, but also appear to be consistent with
a two component distribution.

\end{abstract}

  
\newpage

\section{INTRODUCTION}\label{introduction}

\subsection{Historical background} \label{history}

The issue of whether galaxy disks are significantly opaque is a lively
topic of current research and debate (e.g., Davies \&
Burstein 1995, and papers therein), 
and the conclusions of different workers often
differ widely.  Statistical evidence that most large
presented by Valentijn (1990) based on a diameter-selected 
sample of galaxies from
the ESO survey. Valentijn's result indicated that the bright galaxies
are substantially opaque right out to the 25 $B$ mag arcsec$^{-2}$
isophote.  This conclusion
has been challenged by Burstein and others, who initially questioned
the selection effects inherent in the sample used by Valentijn, and
have maintained that very little extinction is present.  More detailed
studies on individual galaxies have generally concluded in favor of at
least some extinction (e.g., studies of multicolor optical and infrared
images by Block et al.\ 1994, and Peletier et al.\ 1994). There is also
some indication that even galaxy halos may contain dust; Zaritski
(1994) compared faint galaxies in fields bordering nearby spirals
(NGC 2835 and NGC 3521) and in control fields further away on the sky,
and found a small difference in the average colors.  However,
from analysis of surface photometry on a sample of 1700 galaxies,
Giovanelli et al.\ (1995) have recently concluded that less luminous
galaxies are completely transparent, and that even the most luminous
spirals suffer from only small amounts of extinction (0.1 mag) at an
isophotal radius of 23.5 mag arcsec$^{-2}$ in $I$.

Tests of the opacity of spiral disks which rely on intrinsic properties
of the galaxies (surface brightness, inclination angle, isophotal
profiles) are strongly dependent on the necessary assumption of just
how the dust is distributed within the galaxy (Witt, Thronson, \&
Capuano 1992; Witt \& Gordon 1996).  
For instance, layered models with a physically thin but
optically thick dust layer behave like optically thin disks (Disney,
Davies, \& Phillips 1989). In the case of edge-on galaxies, surface
photometry can help determine the geometrical dust distribution
(Kylafis \& Bahcall 1987), but in more general cases the results are
inconclusive. The use of stars or star clusters embedded in the galaxy
can only tell us something about the remaining opacity between us and
the star, and is clearly biased to low opacities. HII regions have also
been used; they have the same bias, but in addition those bright enough
to be easily observed may alter the surrounding ISM so drastically
that they are poor test objects (e.g., Allen et al.\ 1997).

A more direct approach has been proposed by White \& Keel (1992), and
followed up by White, Keel, \& Conselice (1996) and Berlind et
al.\ (1997).  They have looked at partially-overlapping galaxy pairs,
the first two studies at optical wavelengths, the last one in both the
optical and the near-infrared. Assuming that the background galaxy
isophotes are smooth and/or symmetric, they use the unobscured part of
the background galaxy to calculate the amount of background light that
is lost in crossing the foreground disk. All three studies have found
higher extinction in the arms than in the interarm regions, and
extinction curves flatter than Galactic, consistent with a patchy dust
distribution in the foreground galaxy.

The use of background illuminating objects has the advantage that no
assumptions need to be made about the relative distribution of dust and
light in the foreground galaxy. 
Also, the overlapping-galaxy method in particular provides results with 
good spatial resolution. However,
it is necessary to assume that the light distribution in the background
galaxy image is smooth and 
centrally symmetric. In addition, there is an uncertain
contribution of scattered light from the background galaxy into the
line of sight.  But perhaps the biggest disadvantage is that there are
only about two dozen galaxy pairs which offer the right geometry for
the method to be applied (White et al.\ 1996).

\subsection{Distant field galaxies as probes} \label{fieldgal}

A potentially more general way to establish the opacity 
of a galaxy is to look for effects on the
counts, colors, and morphologies of distant field galaxies seen through
the disk of the foreground galaxy. 
This conceptually simple approach
unfortunately suffers from an equally simple problem; the general
distribution of disk stars in the foreground galaxy reduces the
contrast on background galaxy images, and foreground star clusters and
HII regions add strongly to the confusion. High sensitivity is required
in order to reach a surface density of background galaxies which is
large enough to derive a statistically significant result. High angular
resolution and color information are required to assist with the
separation and identification of background galaxy images from objects
in the foreground galaxy.

Hubble Space Telescope (HST) Wide-Field Planetary Camera 2 (WFPC2) 
images offer a significant advantage over images from
ground-based telescopes in their greatly improved resolution. This
reduces the perturbation of faint galaxy images by nearby foreground
objects, and facilitates the separation of stellar objects from
galaxies. Color information is also important; a foreground compact HII
region will be blue compared to the image of a distant galaxy, whether
or not there is any amount of obscuration.  Once galaxies have been
identified, their counts and colors can be compared with counts 
and colors of galaxies in
reference fields not covered by a foreground galaxy. 

\subsection{The data analysis problem} \label{datanalysis}

The HST WFPC2 Archive contains many images of background galaxy fields
and of fields covering parts of nearby galaxies. As an example of the
data available for this study, and an illustration of the analysis
problems which must be overcome before quantitative results can be
obtained, we show in the left panel of Figure 1 one WF
frame (WF3, 1\arcmper3 $\times$ 1\arcmper3 in size) 
from a typical reference field
(the Hubble Deep Field)\footnote{Noise has been added
in order to approximate the shorter integration
times used for the exposures of NGC 4536, as well as the higher sky
background at the position of the spiral. The
mosaic is displayed in the same scale as the 
NGC 4536 color image.}, 
and in the right panel a similar-sized portion
of a deep exposure on NGC 4536. We shall describe both of these images
in more detail in following sections of this paper. For the moment it
is sufficient to note that the differences in the counts and colors of
background galaxies visible in the two panels contain the information
we seek on the opacity of the foreground galaxy. Only a few background
galaxies can be discerned visually with any certainty in the right
panel.  However, before we can conclude that the foreground galaxy is
quite opaque, we obviously need to understand quantitatively how the
data analysis methods we use to determine the counts and colors of the
background galaxies are affected by structural features in the
foreground galaxy.

\subsection{The ``Synthetic Field'' Method} \label{synfield}

In this paper we describe a new method, the {\em Synthetic Field
Method}, which permits us to quantify and calibrate the effects of
crowding and confusion and to determine the accuracy with which
statements can be made about the total opacity and reddening of an
average line of sight through the foreground galaxy.  These synthetic
fields are obtained by adding the HDF reference field images into the
foreground galaxy image; the combined image is then processed in the same
way as the original foreground galaxy image alone.  
The procedure is repeated with different values of extinction
applied to the reference field, 
and a plot drawn of the
final number counts in each simulation versus the assumed extinction in
order to ascertain the ``best'' fit to the actual counts of
real background galaxies.
can be estimated
separately from the biases owing to confusion and crowding in the
foreground galaxy field. 
Finally, a comparison between the average colors of, respectively, 
the control HDF galaxies and the real background galaxies 
will allow an estimate of the degree of reddening.

Compared to the foreground galaxy fields we
analyze, the HDF frames are virtually noiseless; synthetic fields
can therefore be constructed by adding suitably attenuated versions of them
directly into the foreground galaxy images, without degrading the
sensitivity of the data.

Figure \ref{histograms} shows 4 cumulative histograms of galaxy counts
in $I_{F814W}$ images which illustrate the steps in the {\em Synthetic
Field Method}:  Histogram {\em a} is for field galaxies in 
a version of the HDF to which noise has been added
in order to approximate the shorter integration
time used for the exposures of NGC 4536, as well as the higher sky
background at the position of the spiral; {\em b} is for HDF ``control''
galaxies in an un-extincted NGC 4536 simulation\footnote{That is
to say a WF field of NGC 4536 into which the WF fields from the
HDF have been added.}; {\em c} is for HDF galaxies 
in a simulation with 
$\sim$ 0.6 mag of extinction at $I$\footnote{The $I$ control field has
been attenuated to mimic $\sim$ 0.6 mag of extinction.};
and finally {\em d} is for real background
galaxies in the NGC 4536 field. From the figure, we would
conclude that there is about 1 mag of extinction at $I$ averaged over
the WF2 and WF3 fields in this region of the foreground galaxy\footnote{The
results for each separate field do not differ significantly; the
numbers of galaxies have been normalized to the area of one
WF field.}, and about 0.1 mag of average extinction at $I$
in the region of NGC 4536 covered by the WF4 field. 

\subsection{Limitations of the method} \label{limitations}

Clearly, one major limitation of our method is that the results are only
statistical in nature, i.e., the method does not give an answer for the
extinction and reddening along any {\em specific} line of sight through the
foreground galaxy, but only a statistical estimate of these quantities
averaged over some substantial fraction of the foreground galaxy
image.  The accuracy of the method is limited not only by
the counting statistics of background galaxies, but also by the
intrinsic inhomogeneities in their spatial distribution.
These uncertainties must also be quantified in
order to provide a convincing result. In addition, there are several
smaller selection effects which plague the analysis and which will be
discussed later; for instance, the fact that field galaxies
also show a mild color--magnitude dependence 
(Tyson 1988; Williams et al.\ 1996)
must also be taken into account.

Another limitation is that, while the 
numbers of galaxies probe the whole disk 
(there is useful information in both detecting or not detecting a 
background galaxy),
the conclusions from the colors 
of the galaxies will be biased in favor of
regions of lower opacity (because we need to detect the 
galaxies in order to measure there colors). 
Therefore, it is possible to determine an extinction law by combining
the number-count and color information only when we can safely
assume that nowhere in the region of interest is
the extinction so high that we
cannot sample it with background galaxies (see Appendix). 

\subsection{Plan for this paper} \label{plan}

In this paper we present a description of the {\em Synthetic Field Method}
through a detailed application to deep WFPC2 images of the Sc spiral
NGC 4536. In order to examine the range of applicability of the method,
we also analyze shallow images of the SBm Magellanic dwarf irregular
NGC 3664 (DDO95). We find substantial opacity in both galaxies,
although the precision of the result is low for NGC 3664, a consequence
of the shorter exposure time used for the observations of this galaxy. 
In future
papers we intend to apply the method to images of other nearby galaxies
drawn from the HST archives.

We begin in \S \ref{reffield} with a discussion of the reference field
and the processing necessary in order to render it useful for our
present purposes.  Next in \S \ref{N4536} we describe the data on the
foregound galaxy NGC 4536 which we have chosen to illustrate the
method. Section \ref{photometry} describes the methods we use to
identify and measure the magnitudes of field galaxies. Section
\ref{results} presents the synthetic field simulations and the results on
extinction through the disk of NGC 4536. 
Section \ref{N3664} presents an abbreviated discussion
applying the method to NGC 3664. 
In \S \ref{errors} we 
examine the various sources of errors and biases we have been able to
identify.
Finally, we summarize and discuss
our results on the extinction and reddening through these
two galaxies in \S \ref{summary}.

\section{THE REFERENCE FIELD} \label{reffield}

The choice of the reference field is clearly of prime importance to
this method. For simplicity, images of this field ought to be available
with the same camera and range of colors as the images of the
foreground galaxy, and ought to be entirely representative of the sky
in areas free of large nearby foreground galaxies. There are by now
many fields observed with the WFPC2 which could satisfy these criteria,
including the Medium Deep Survey fields (typically exposures of
\gax 5000 s in the F606W and F814W passbands;
Griffiths et al.\ 1994; Ratnatunga, Griffiths, \& Ostrander 1998) and the
field around the weak radio-galaxy 53W002
(24 orbits in 3 colors; e.g., Pascarelle et
al.\ 1996a,b).  However, the {\em Hubble
Deep Field} (HDF; 150 orbits in 4 colors, Williams et al.\ 1996) offers
the deepest set of images. Its properties are also
presently the object of intense scrutiny by many workers, and detailed
analyses of the counts, colors, and redshifts of the
galaxies visible in it are becoming available
(e.g., Cohen et al.\ 1996; Gwyn \& Hartwick 1996; 
Villumsen, Freudling, \& Da Costa 1997; Colley et al.\ 1996;
Madau et al.\ 1996).  
The question as to what
extent our chosen reference field is ``representative'' can therefore
be answered with considerable certainty.  For example, at $m_{F814W,AB}
\sim$ 26 (appropriate to the present work), the number-counts of the
HDF galaxies are known to be complete, and the galaxy 
number-magnitude relation is similar to that of
other fields (Williams et al.\ 1996). The HDF therefore
provides an excellent reference field for our purposes.

\subsection{Processing of the HDF data} \label{HDFdata}

We used the Version 2 Release WFC images of the HDF (Williams et al.\ 
1996) retrieved from the STScI/HST archive. This release of the HDF was
made from dithered exposures using the ``drizzling'' method (Fruchter \&
Hook 1997).  There are 3 contiguous WFC fields in this data set (and
one PC field which we do not use here). Each WFC field was recorded in
the 4 photometric bands designated F300W, F450W, F606W, and F814W
(cf.\ the HST Data Handbook); we have not used the images in the UV
band F336W.

Several processing steps were carried out on the HDF data set in order
to render it compatible with the images of NGC 4536 and NGC 3664.
First, the HDF images were all re-binned to their original pixel size of
0\arcsper1 since this is the scale of the (un-dithered) exposures of
NGC 4536 and NGC 3664 (and most other exposures of nearby galaxies in
the HST archives).

Second, since there is no HDF image in the F555W passband
used for the foreground galaxies, we created one by interpolating
linearly between the existing F450W and F606W images, taking into
account the differing responses of the three filters.  This procedure
will produce the right result for galaxies that have a linear spectrum
between 4520 ${\rm \AA}$ and 5935 ${\rm \AA}$, and an error in the flux
at 5400 ${\rm \AA}$ proportional to the deviation of a galaxy spectrum
from linear in this wavelength range.  For redshifts lower than $z$ =
0.13 (v $\sim$  40000 km~s$^{-1}$), the 4000 ${\rm \AA}$ break is
blueward of the interpolation range, and the spectra should be fairly
linear.  However, a significant fraction of the HDF galaxies recovered
in the simulations have $0.13 < z < 1$.  Most of the recovered HDF
galaxies have an observed (isophotal) $m_{F814W}$ brighter than 25
(Figure \ref{fhdfi}). From the estimates of Villumsen 
et al.\ (1997) for $R$ magnitude limits, and assuming $(R - I) \sim 0.5$
(Smail et al.\ 1995), the recovered HDF galaxies should be at
redshifts $z \le 1$.  These galaxies are also mostly brighter than
$m_{F555W,{\rm interpolated}}$ = 26 (Figure \ref{fhdfv}).  The majority
of HDF galaxies brighter than $m_{F606W}$ = 26 are bluer than 1.5 in
($B_{F450W} - V_{F606W}$) (Williams et al.\ 1996, their Figure 32). These
estimates allow us to model the error in our interpolated F555W image
by using stellar population synthesis models (Babul \& Ferguson 1996)
in order to produce spectra of galaxies with different star formation rates.
From the results for the synthetic spectra, galaxies with $z<$ 1 and
($B_{F450W} - V_{F606W}$) $<$ 1.5 could suffer from a systematic shift
of $\sim$ -0.03 mag in $m_{F555W,{\rm interpolated}}$.  The error will
be in the sense of {\em very} slightly overestimating the reddening of
the real galaxies relative to the simulations.
However, the
dispersion in ($m_{F555W,{\rm interpolated}} - m_{F555W,{\rm real}}$)
is $\sim$ 0.08 mag, of the same order as the dispersion of the mean
($m_{F555W,{\rm interpolated}} - m_{F814W}$) color between the 3 HDF
WFC fields. We conclude that the error introduced into the interpolated
F555W image is negligible.

\subsection{Making synthetic fields with the HDF} \label{synthetic}

The HDF images at the full sensitivity of the original observations are
added directly to the images of the foreground galaxies in order to
create the synthetic fields. The signal-to-noise 
ratio of the HDF is so high that this
simple addition results in an image which retains the noise level of
the foreground galaxy image. Before the addition, the HDF can be
attenuated and reddened by any amount in order to create a set of
synthetic images in which one can search for and measure ``real''
background galaxies as well as the HDF ``control'' galaxies.

In order to 
provide a visual demonstration of how many galaxies in a
``representative'' field are lost to {\em both} crowding
{\em and} extinction,
we have also created a special version of the HDF,
simply by adding an appropriate amount of
noise to it. We refer to this as
the ``degraded'' HDF.
After adding the noise,
its signal-to-noise ratio is the same as 
that of the images of
the foreground galaxy, which have a shorter exposure time and
a higher sky background.
The WF3 frame from the ``degraded'' HDF image
is shown in Figure 1 (left panel). We produced this color
image by combining together the $V_{F555W,{\rm interpolated}}$ frame
as blue, the $I_{F814W}$ frame as red, and an average of the two as
green. The image is displayed in a square root scale which favors the
dynamic range of the NGC 4536 color image; this latter image is shown
in the right panel of Figure 1. 
The ``degraded'' HDF represents the number of galaxies that
would be recovered in the {\em total absence of any foreground galaxy},
and hence without confusion with globular clusters, HII regions, and
stars, but
at the same position in the sky and with the same exposure time
as the images of NGC 4536.

The differences in the
numbers of background gal- axies which can be recognized between the two
frames in Figure 1 is largely due to crowding and   
confusion, as will be seen in\ \S\ref{completeness} from the comparison 
between the ``degraded'' HDF and 
the synthetic images without extinction\footnote{Synthetic fields produced
by adding the HDF with zero attenuation to the NGC 4536 frames.}. 
These data, however, also imply a significant
amount of extinction, as we shall subsequently show.
The power of the synthetic field method resides in its ability
to separate the effects of extinction from those of confusion and
crowding; such a large number of galaxies is lost to crowding only,
and it has such significant effects on the photometry (\S\ref{photerr}),
that it would be impossible to do this project without accounting
properly for it.

\section{DEEP IMAGES OF NGC 4536} \label{N4536}

\subsection{Observations} \label{obs4536}

The images of NGC 4536 were originally acquired with the Hubble Space
Telescope (HST) Wide-Field Planetary Camera 2 (WFPC2; Holtzman et al.\ 
1995) for the Distance Scale Key Project; they have already been
published by Saha et al.\ (1996), who give a very detailed account of
the observations and the pipeline calibration of the data. The WFPC2
field is shown in Figure \ref{fn4536}, overlaid on an STScI Digitized
Sky Survey image of the galaxy. We do not use the PC frame (WF1)
in this work.

NGC 4536 is a bright spiral galaxy Sbc(rs)I-II (de Vaucouleurs et al.\ 
1991), with $R_{25}$ = 3\arcmper79, in the direction of Virgo, at a
distance of 16.2 Mpc (Saha et al.\ 1996). This yields a spatial
resolution of $\sim$ 8 pc per WF pixel. The Galactic extinction
towards NGC 4536 is $A_B$ = 0.00 (de Vaucouleurs et al.\ 1991).  

There are 34 individual exposures in the F555W passband, taken at 17
discrete epochs between 1994 June 3 and 1994 August 9; in the F814W
passband, 10 images were taken at 5 discrete epochs, over the same
two-month period.  For each filter, two individual exposures 2000
seconds long were taken back-to-back in each epoch, to allow for cosmic
ray (CR) removal by an anti-coincidence technique described in
Anderson et al.\ 1995 and in Saha et al.\ 1996.

\subsection{Further processing} \label{proc4536}

We began working on the images as retrieved from the archive, i.e.,
bias-subtracted and flat-fielded.  The values of ``warm'' pixels
(identified in the calibration pipeline as having a
moderately high dark count during the time of observation) were repaired
by replacing the measured dark current with the interpolated dark value
between two bracketing dark frames (Voit 1997). We then removed the
cosmic rays in each image using the ``CR-split'' procedure (Anderson et al.\ 
1995).

Since we are not looking for time variability, we combined all the data
in each passband into a single deep image.  Relative shifts between
the different epochs were found with a cross-correlation technique.
The images at all epochs were nominally taken with the
same telescope roll-angle; however, given that distinguishing between
point and extended objects is crucial for our project, we also
used this technique to check that in fact there were no unintended
relative rotations between the frames.  Finally, the images were
corrected for geometric distortion (Trauger et al.\ 1995), aligned, and
co-added using the ``Drizzle'' variable-pixel linear reconstruction algorithm
(Fruchter \& Hook 1997).  The resulting images have
total exposure times of, respectively, $68 \times 10^3$ seconds in the
F555W passband, and $20 \times 10^3$ seconds in the F814W passband.

\section{BACKGROUND GALAXY\\ PHOTOMETRY} \label{photometry}

\subsection{Object extraction} \label{sextractor}

We used SExtractor (Bertin \& Arnouts 1996) to perform the $I_{F814W}$
and $V_{F555W}$ photometry on all objects, up to the $24.5\ {\rm
mag}_{AB}$ arcsec$^{-2}$ isophote in $I_{F814W}$.  
The parameters for the search and
photometry were chosen to optimize the background determination, as
judged from the lack of ``holes'' (oversubtracted regions) in the 
background-subtracted NGC 4536 images. 
Likewise, the threshold surface brightness
was selected so as not to ``find'' noise. We performed the
search and fixed the threshold in the
F814W passband not only because of the actual selection bias in favor
of red objects (see below \S\ref{colbias}), 
but also because our photometry was limited
by the shorter exposure time of the $I_{F814W}$ frame. 

From the list of all objects found by SExtractor, we selected
the background galaxies by visual inspection of the images. 
We used the morphology, the fuzziness and, especially in 
crowded regions, the color. In crowded regions, we selected
preferentially fuzzy objects with $(V_{F555W} - I_{F814W}) \ge 0.5$. 
Our search and selection criteria mean that, in principle, 
from the number-counts we will measure the extinction at $I_{F814W}$. 

Figure \ref{cbgi4536} 
shows the cumulative histograms
of real galaxies found in the background of NGC 4536, in the 
F814W passband.
The left panel of this figure
displays the results for the WF2 and WF3 fields combined (the numbers 
are normalized 
to the area of one WF field); the right panel shows the results for the
WF4 image.  We analyzed the WF2 and WF3 images of NGC 4536 together,
given that they span most of the northwestern arm of the galaxy. In
contrast, the WF4 field covers the northern interarm region and a small
area that appears to be beyond the visible disk of the galaxy.
We found a total of, respectively,
14 and 16 real background galaxies in the
WF2 and WF3 fields, and 35 galaxies in the WF4 field.
The difference between the number-counts in this figure
and the number-counts of the HDF galaxies (Figure \ref{cumi})
is caused by the combined effects of systematics,
crowding and, finally, what we are interested in: extinction.

\section{SIMULATIONS AND RESULTS\\ FOR NGC 4536} \label{results}

Synthetic \ \ fields \ \ were \ \ produced \ \ with \ varying \\amounts of (spatially
uniform) attenuation and
reddening (Table 2).  The idea is to attenuate the HDF
fields until one recovers the same number of simulated and real
background galaxies. This attenuation,
derived from the number-counts,
should correspond to the average extinction through the disk of NGC
4536 to an uncertainty given by the Poisson and clustering errors.  
In practice, what we measure is:

\begin{equation}
A_I = -2.5 \cdot C \cdot log(\frac{N}{N_0}).
\end{equation}

\noindent $N_0$ is the normalization, in this case the number of
HDF galaxies recovered in the synthetic fields without attenuation,
and $C$ parametrizes the dependence of the number vs. extinction
curve on object selection biases (\S\ref{ext4536}). 

The reddening
suffered by the background galaxies is found by comparing their average
color to the average color of the HDF galaxies recovered from the
synthetic fields. Because galaxies have a color--magnitude dependence,
one must be careful to perform this comparison with simulations that
have the same extinction as that measured (from the counts) for the
real galaxies or that have the same magnitude limit as the
extinction-corrected limit of the real galaxies.
The selection biases of the real and synthetic samples of
galaxies, and especially the color selection bias
(\S\ref{colbias}), should also be the same for the comparison to
be meaningful. 
 
To produce the synthetic fields, we: (1) scaled the HDF WFC frames 
to the exposure times of the NGC 4536 mosaics; (2) attenuated the
HDF frames to mimic different amounts of attenuation and reddening; 
(3) added them in directly to the images of the spiral.
To improve the statistics, at each attenuation/reddening 
we added the three HDF WFC frames, one at a time, 
to each one of the NGC 4536 WFC frames. 
What we call one simulation includes the results
of repeating the procedure with 3 synthetic fields, one for each
HDF WFC frames\footnote{In all the figures, the numbers of recovered
HDF galaxies
are averages of the 3 synthetic fields.}.
As in the case of the real background galaxies, 
the photometry on the synthetic frames
was done with SExtractor, and the threshold surface brightness was
fixed in the $I$ image.
The identification of control galaxies was
done, like that of real background galaxies, by visual inspection
of each synthetic frame
(as opposed to, for example, through cross-correlation with a
list of known HDF galaxies). Systematics and confusion should thereby
affect synthetic and real galaxies in the same way.

Since we did not know {\em a priori} the extinction law for the disk,
we made simulations with both Galactic reddening (Fitzpatrick 1986) and
completely grey extinction (i.e., the HDF was attenuated equally in
both the F555W and the F814W passbands).  We hypothesize that the
extinction through galaxy disks will lie somewhere in between these two
limiting cases.  Another reason to have the two sets of simulations is that
they should establish a
relative scale that takes into account all possible systematic errors,
against which to measure the attenuation and the reddening of the
background galaxies.  We also constructed synthetic field simulations
with a ``mixed'' extinction law (the HDF $V_{F555W}$ frame was more
attenuated than the $I_{F814W}$ frame, but less than for Galactic
reddening), to check that the color scale established by the two limiting
cases was roughly linear.  We covered a large range in attenuation,
from no attenuation at all up to 2.6 mag in both passbands. Since NGC
4536 is inclined at $65\deg$ to the line of sight (de Vaucouleurs et
al.\ 1991), we performed these highly-attenuated simulations
anticipating the possibility of a face-on optical depth of $\tau_{V,
Johnson}$ = 1, both with grey and with Galactic reddening extinction
curves.  Table 2 shows the input parameters of the
simulations, as well as the number and average output color of the HDF
galaxies recovered from the NGC 4536 frames; a line indicating the 
actual NGC 4536 observations is included for comparison.

\subsection{Completeness} \label{completeness}

To assess the effects of crowding alone, 
we compared the ``degraded''
HDF frames (\S\ref{synthetic}) to the simulation without extinction, i.e.,
a synthetic field where the HDF was added to the NGC 4536 frames
without any attenuation.  Figures \ref{ihist} and \ref{vhist} compare
the average numbers, of HDF galaxies only, recovered from the NGC 4536
frames in the simulation, against the average number per $m_{F814W}$
and $m_{F555W}$ bin, respectively, of galaxies in the ``degraded'' HDF.
The effect of crowding on the number-counts is, indeed,
dramatic.

Given the isophotal $I_{F814W}$ brightness limit, while the HDF
galaxies in the ``degraded'' frame are complete 
to at least $m_{F814W}$ = 25, when added to the
NGC 4536 they are complete only to $m_{F814W}$ = 24 in the WF4 frame
and, optimistically, to $m_{F814W}$ = 23 in the WF2 and WF3 frames.
This is the range where galaxy clustering is most noticeable
(\S\ref{clusterr}).  On the other hand, we were able to detect
galaxies up to an integrated (isophotal) magnitude of $m_{F814W}$ = 26
in all frames.  The $V_{F555W}$ histogram shows that we were able to
detect many more blue, faint galaxies in the WF4 NGC 4536 field.
 
The histograms of the recovered synthetic galaxies, however, might
suffer from two competing effects:  the confusion of faint wings with
the foreground disk, and the contamination of the photometric aperture
with foreground objects.  As described below (\S\ref{bgerr}), when
using isophotal magnitudes galaxies are measured, on average, 0.1 mag
too bright in the F814W bandpass, and up to 0.6 mag in the F555W
filter, depending both on the crowding and on the faintness of the
object.  Fortunately, our investigation does not require absolute or
total measurements of the galaxy brightnesses, as long as everything is
compared in the same {\em relative} scale provided by the simulations.
 
An absolute measurement of the total magnitude limit is needed only to
estimate the expected clustering error.  For this purpose, we correct
the $m_{F814W}$ completeness limit for the faulty background
subtraction, and we also assume that total magnitudes are 0.3 brighter
than isophotal. After an additional correction for the extinction
(\S\ref{ext4536}), we set our completeness limits at $m_{F814W}$ = 23.7
in the WF4 field, and at $m_{F814W}$ = 21.7 in the WF2 and WF3 frames.

\subsection{Real vs. control galaxies}
\label{com4536}

Figures \ref{simreali}, \ref{simrealv}, and \ref{simrealcol} compare
the number of real galaxies to the number of HDF ``control'' 
galaxies recovered in
the simulations without attenuation, per $I_{F814W}$ (isophotal)
magnitude bin, $V_{F555W}$ (isophotal) magnitude bin, and color bin,
respectively.  The effects of extinction are readily noticeable in the
WF2 and WF3 fields, both from the reduction in the number-counts and
from the shift to fainter magnitudes of the real galaxies. For the WF4
field, any effects are much less apparent. The reddening, though, is
harder to discern from Figure \ref{simrealcol}, even for the more
extincted WF2 and WF3 images.

The relative photometric error for the individual galaxies was
estimated by comparing the three measurements we had of each one of
the simulated objects in the synthetic fields 
without attenuation (each HDF frame was added to
each NGC 4536 frame). Therefore, it includes the effect of differential
crowding, and amounts to 0.16 mag at $I_{F814W}$ (isophotal) and 0.26
mag at $V_{F555W}$ (isophotal). However, the error in ($V_{F555W} -
I_{F814W}$) for each galaxy is only 0.10 mag, given that the color has
been measured using a small, fixed aperture of 0\arcsper5 diameter,
precisely to minimize the impact of crowding (see~ \S\ref{bgerr}).

\subsection{The extinction measurement}
\label{ext4536}

Figure \ref{fext4536} shows the comparison between the number of real
galaxies (horizontal lines) and the average number of 
``control'' galaxies recovered in the
simulations at each given $I_{F814W}$ attenuation.  
The top panel
displays the results for the WF2 and WF3 fields;  
the bottom panel, those for the WF4 image.
The solid triangles are for simulations with Galactic reddening;
the open triangles, for synthetic fields with grey extinction.
The different shapes of the curves for both types of extinction
illustrate the dependence of the number-counts 
on the colors of the galaxies. 
The actual measurement
of the extinction $A_I$ was performed by linearly interpolating between
the simulations bracketing the real galaxies.  The error bars include
Poisson and clustering errors (\S\ref{clusterr}).

We find for the arm region of NGC 4536 (WF2 and WF3 fields) an
absorption of $A_I = 0.74 \pm 0.49$ mag in the case of grey extinction,
and $A_I = 1.07 \pm 0.49$ mag in the case of a Galactic reddening law.
Without the contribution for clustering, the purely Poisson error is
$\pm 0.26$ mag.  For the interarm and outside area (WF4), the results
are, respectively, $A_I < 0.46$ mag ($< 0.33$ mag without clustering
error) for grey extinction, and $A_I <$ 0.52 mag ($< 0.39$ mag without
clustering error) for Galactic reddening. The $+1 \sigma$ error of 0.37
(0.21) mag is already included; the figure shows that the difference
between the attenuation inferred from the two models increases with
extinction. The difference is in the sense that we expect; it is due to
our color selection bias, which favors the detection of faint galaxies
(statistically blue) when they are reddened (\S\ref{colbias}). The
difference, however, is smaller than the errors.

\subsection{The reddening measurement}
\label{red4536}

For the background galaxies, we measure a mean color of
$<{\rm (V - I)}>_{WF2\&3} = 1.36 \pm 0.14$ in the arm, and of
$<{\rm (V - I)}>_{WF4} = 1.40 \pm 0.12$ in the interarm
and ``outside'' region.
Figure \ref{fred4536} shows the comparison of the color 
of the background galaxies to those of the simulations. We remind the
reader that this is the appropriate comparison, rather than a
confrontation between the two samples (i.e., WF2+WF3 and WF4), given
their different magnitude limits and all the systematics that come into
play in the presence of severe crowding.  
We also point out that the purpose of this figure is to
compare the color of the background galaxies to the synthetic
fields closest in extinction and object selection biases;
the locus of the real galaxies in the plot does not necessarily
imply an extinction law for the obscuring material in the
foreground disk, unless we can safely assume that 
the whole region is sampled  
by galaxies whose colors we can measure 
(\S\ref{discussion} and Appendix).
The left panel shows the
results for the WF2 and WF3 fields; the right panel, those for WF4.
The simulations with Galactic reddening (open triangles) do not follow
exactly the theoretical extinction law (solid line) for large
absorptions, neither do the simulations with grey extinction (filled
triangles) follow a constant line in color.  The departure of the grey
simulations from the zero-reddening line is due to our selection for
red galaxies ~\S\ref{colbias}, 
and to the color--magnitude relation of galaxies.  As we
go to larger extinctions, given our detection magnitude limit we are
left with statistically brighter galaxies, which are also intrinsically
redder.  The deviation of the reddened simulations from the theoretical
Galactic reddening line is due to the systematic error in the
background subtraction in crowded areas described in ~\S\ref{bgerr}.
In the case of WF4, even if it is less crowded, we still see the effect
because our magnitude limit is fainter, and faint galaxies are
proportionally more affected by the problem.  This is also the cause of
the relative shift between the scales of the two panels. The WF2 and
WF3 images of NGC 4536 are more crowded than the WF4 field;
statistically, {\em all} galaxies in the former fields have been
measured relatively bluer, producing the downward shift of the whole
color vs. extinction plot.
All these systematic effects will affect the real galaxies in the same
way and should not bias the estimate of the reddening,
as long as the
scale defined by the simulations with the two limiting extinction laws
is linear. 
To check that this is the case we have performed
simulations with a ``mixed'' extinction
law\footnote{That is, the $V$ light of each individual galaxy is more attenuated
than the $I$ light, but less than in the Galactic reddening case;
this type of extinction can be caused by an unresolved
clumpy dust distribution (Witt \& Gordon 1996)}. 
The output of the ``mixed'' simulations 
is shown in Figure \ref{fmix4536}.
When compared with their input line, the simulations
with a ``mixed'' reddening law seem more affected by the selection bias
that pushes the grey simulations redward than by the background
subtraction that pushes the Galactic reddening simulations blueward.
The simulations, however, follow the input line within the errors,
which shows that the color scale is, indeed, roughly linear. 

The background galaxies behind the northwestern arm of NGC 4536
(WF2 and WF3 fields, Figure \ref{fred4536}, left)
fall in between the lines for the completely grey extinction and 
the Galactic reddening law in the
the color vs. extinction plot.
Since, unfortunately, the color line of the grey simulations shows a kink
precisely at the location of the real background galaxies, to find the
reddening we will rather use the color obtained by interpolating
between the grey simulations with $A_I$ = 0.58 mag and $A_I$ = 1.51
mag, at the extinctions (inferred from the grey and Galactic reddening
models, respectively) of the background  galaxies.  The interpolation
yields
$<$($V - I$)$> = 1.22 \pm 0.11$ at $A_I$ = 0.74 mag, and 
$<$($V - I$)$> = 1.25 \pm 0.11$ at $A_I$ = 1.07 mag. 
Therefore, the background galaxies suffer a reddening of $E(V_{F555W} -
I_{F814W}) < 0.32$ mag (grey extinction model) or $E(V_{F555W} -
I_{F814W}) < 0.29$ mag (Galactic reddening law). The error in the
reddening, already included, is 0.18 mag.

The background galaxies in the interarm region (WF4,
Figure \ref{fred4536}, right), however, seem to follow the Galactic
reddening line.  Comparing their color to the mean color of the
simulation without extinction, they exhibit a reddening of $E(V_{F555W}
- I_{F814W}) = 0.20 \pm 0.13$ mag, regardless of the attenuation model
used.

\section{NGC 3664} \label{N3664}

\subsection {Observations and Data Reduction.}
\label{d3664}

NGC 3664 was observed with the WFPC2 in 
1996 March 3.
The total exposure times are 600 seconds in the
F555W passband, 800 seconds in the F814W passband, 
and 1600 seconds in the
F439W passband.
NGC 3664 is a Magellanic peculiar (SBm(s)IV-V, de Vaucouleurs et
al. 1991). From the recession velocities ($v_{GSR}$) cited in this
reference for NGC 3664 and NGC 4536, and using the distance to
NGC 4536 above (\S\ref{obs4536}), NGC 3664 is at a distance of 11.9 Mpc. 
The data will then have
a spatial resolution of $\sim$ 6 pc per WF pixel.
We assume the foreground Galactic absorption is $A_B$ = 0.12 mag
(de Vaucouleurs et al.\ 1991).

Figure \ref{fn3664} shows the WFPC2 field,
overlaid on an STScI Digitized Sky Survey image of the galaxy.
Once again, we do not use the PC (WF1) frame.
NGC 3664 covers only 45\% of the three WFC chips; this decreases the
number of galaxies expected to be seen through the disk, 
but provides a fairly large surrounding field for
additional comparison.  From its morphology, NGC 3664 is believed to be
seen face-on.  

Partly because of the shallowness of the data, partly because
the work on NGC 3664 started out in a relatively independent way,
and also owing to a desire to explore diverse lines of
attack, the analysis of the NGC 3664 data differs in the details from
that of NGC 4536.  The main differences are: (1) the subtraction of
the available (and properly scaled) $B_{F439W}$ image from the $I$ image   
to facilitate the search for relatively red objects (\S\ref{f3664});
(2) the use of the Rieke \& Lebofsky (1985) Galactic reddening curve;
(3) the semi-automatic identification of background galaxies, 
both real and simulated (\S\ref{hdf3664}). To improve the statistics,
we did more simulations at smaller extinction intervals (Table 
3). 
We did not try to fix the warm pixels in the raw images, and cosmic
rays were iteratively removed based on the deviation of
individual pixel values from the mean.
Finally, we adjusted the parameters of SExtractor 
to minimize the scatter of the difference
between the measured brightnesses of the
HDF galaxies before and after adding them
to the NGC 3664 fields. For the same reason,
in what follows we use the ``automatic'' aperture
magnitudes calculated by SExtractor, rather
than the isophotal magnitudes\footnote{``Automatic'' apertures are
a constant, in this case 2.5, times the size of the Kron
``radius'', i.e.,
an elliptical aperture that contains the same fraction
of the total light for all objects, regardless of
their brightness profile.}.

\subsection{Finding background galaxies}
\label{f3664}

To remove most of the bright, blue star-forming regions
in the disk of NGC 3664, 
we subtracted a properly-scaled 
F439W image from the F814W image.
The scaling was done arbitrarily, 
to remove this blue component without creating substantial 
``holes'' in the residual image. This method is reasonable, 
because we cannot or can only barely see the background
galaxies in 
the F439W passband. 
The disadvantage of this procedure is that it increases the noise. However,  
since background galaxies are identified primarily through
their relatively red appearance, the disadvantage
is far outweighed by removing the 
bright, relatively blue, foreground in the first step.
The search for galaxies with SExtractor was performed in this artificial, 
blue-subtracted image. The photometry of the identified 
galaxies, however, was done
in the original F555W and F814W frames. 

\subsection{The synthetic field simulations}
\label{hdf3664}

Table 3 shows the input parameters of 
the simulations, and the number and average output color of the HDF 
galaxies recovered behind and surrounding
NGC 3664; a line indicating the 
actual NGC 3664 observations is included for comparison.
As opposed to what we did in the case of NGC 4536, the selection
of galaxies out of the object catalogs produced by SExtractor
was done in a semi-automatic fashion.
For example, we only selected objects redder 
than ($V_{F555W} - I_{F814W}$) = 0.17 mag
(in the catalog, i.e., several tenths of mag redder in
actuality in crowded regions, see ~\S\ref{bgerr}).
Other selection criteria were a 
(SExtractor) star-galaxy separation parameter
``class\_star'' $<$ 0.8
(the values go from 0.0 for an unmistakable galaxy to 1.0 for
a perfect star); a FWHM of a Gaussian fit to the  
object profile $>$ 0\arcsper2; and an error in the 
``automatic'' aperture
magnitude  $<$ 0.17 mag.
In bright foreground star forming regions, the search 
and measurement of background
galaxies is virtually impossible. Therefore, we excluded 
from the search regions of NGC 3664 with a surface brightness
higher than 21\ mag arcsec$^{-2}$ (WF3) and 20.7\ mag arcsec$^{-2}$
(WF4), respectively\footnote{These limits were set empirically, with the
aid of the simulations, so as to retrieve the maximum number of 
HDF galaxies and the minimum amount of misidentifications}. 
We still had to exclude by hand saturated stars that were included
in the automatic selection of galaxies. 

\subsection{Real vs. control galaxies}
\label{com3664}
 
Because of the difficulties in identifying galaxies reliably,
we decided to exclude 
from the sample behind the disk of NGC 3664
all candidates fainter than $m_{F814W}$ = 22.8.
This reduced our sample of real galaxies found behind
the projected disk of NGC 3664 to 8 objects.
To the same limiting magnitude of $m_{F814W}$ = 22.8,
in the simulation without extinction
we recovered an average of 17 HDF galaxies within the disk.
The discrepancy must be caused by extinction.
For the areas outside the visible disk of the galaxy,
we kept galaxies up to an integrated magnitude
of $m_{F814W}$ = 24. We found a total of 38 
background galaxies in the
surrounding field, vs. an average of 37 control galaxies.

\subsection{Extinction in NGC 3664}
\label{ext3664}

We find for the disk of NGC 3664 
(Figure \ref{fext3664}, {\em top})
an absorption of $A_I = 1.01 \pm 0.63$ mag 
in the case of grey absorption, and
$A_I = 1.02 \pm 0.63$ mag in the case of a Galactic reddening law. 
The purely Poisson error is $\pm 0.49$ mag.
For the surrounding area 
(Figure \ref{fext3664}, {\em bottom}) 
the results are, respectively, 
$A_I <$ 0.31 mag ($<$ 0.17 mag without clustering error) 
for grey extinction and  $A_I <$ 0.30 mag 
($<$ 0.16 mag without clustering error)
for Galactic reddening law.
(The results have been corrected for foreground extinction
in the Milky Way.)
Table 1 lists the relevant parameters
for the calculation of the error in the extinction.

\subsection{Reddening in NGC 3664}
\label{red3664}
 
For the background galaxies, we measure a mean color
of $<$($V - I$)$>_{inside} = 1.46 \pm 0.19$ behind the
disk of NGC 3664, and of $<$($V - I$)$>_{outside} = 1.13 \pm 0.10$
in the surrounding field.
Figure \ref{fred3664} shows how we compare the 
color and extinction of the background galaxies to
the simulations.
The left panel exhibits the results for the
field covered by the disk of NGC 3664. 

As in the case of NGC 4536, 
the background galaxies behind the disk of NGC 3664 lie in between the lines
for the grey and Galactic reddening simulations.
To find the reddening, we compare the mean color of the background 
galaxies to the mean color of the grey simulation with the
closest amount of extinction, i.e., the one with $A_I$ = $A_V$ = 0.95 mag.
The output mean color of this simulation is  
$<$($V - I$)$> = 1.06 \pm 0.11$. Therefore, the background
galaxies suffer a reddening of 
$E(V_{F555W} - I_{F814W}) = 0.40 \pm 0.22$ mag.

The right panel of Figure \ref{fred3664}
shows the results for the galaxies recovered
outside the visible disk of NGC 3664. 
The dashed line 
follows the color of the grey
simulations after correcting 
for incompleteness: 
a shift was added to the average color of the
recovered galaxies as a function of increasing
extinction, so that it would follow the same 
behavior as the average color of the (un-added) HDF galaxies 
with a progressively brighter magnitude limit.
Although we prefer to use the color without correction,
because otherwise the small numbers make the error 
in the color very large, the line does indicate that
we are seeing an incompleteness effect. Conversely,
in the case of the simulations with Galactic reddening,
the line after correction (not plotted) basically follows
the output mean color of the recovered galaxies. This
confirms that, given our color selection bias, 
we can still find relatively faint galaxies, even in the presence
of extinction, as long as they are reddened. 
The result for the 
background galaxies outside the visible disk of
NGC 3664 is that their number and color 
are consistent with no attenuation.

\section{SYSTEMATIC ERRORS AND BIASES} \label{errors}

\subsection{Color selection and the color--magnitude relation}
\label{colbias}

Because spiral disks are relatively blue, red background galaxies are
more readily detected behind a foreground disk.  As mentioned above, in
crowded regions of NGC 4536 we selected in practice for red galaxies
[i.e., with $(V_{F555W} - I_{F814W}) \ge 0.5$], to be able to
distinguish them from the stellar content and HII regions (which are
predominantly bluer) in the foreground galaxy.  This bias will be even
more noticeable in grey simulations, where (given a certain attenuation
at F555W) the F814W passband will be more attenuated than in the
corresponding simulation with Galactic reddening.  As a result, more
faint blue galaxies will be discarded in the grey simulation than in
the reddened one, the mean color measured for the outcome of the grey
simulation will be redder than otherwise, and the measured color shift
between it and a reddened simulation will be smaller (bluer) than the
input.  The color--magnitude relation of galaxies (intrinsically
brighter galaxies tend to be redder) will have an effect in the same
direction:  at higher extinctions, the statistically brighter galaxies
will be detected, and the average color of the sample will be redder.

\subsection{Systematic error in the color \\measurement
due to the background}
\label{bgerr}

As it searches for and measures objects, SExtractor determines the
background with a sigma-clipping algorithm, plus a modified mode in
crowded areas. In very crowded fields, however, this results in an
underestimation of the actual background when using large apertures
(larger than 0\arcsper6 in radius for the two galaxies investigated
here).  The background determination algorithm has the purpose,
precisely, of removing bright spots (stars, clusters or HII regions)
from the measurement of the background, so that faint, extended objects
can be identified. On the other hand, the larger the photometric
aperture, the higher the probability that these same bright spots in
the foreground galaxy disk will lie within the aperture, and will be
included in the measurement of the flux of the object of interest, in
this case, the background galaxy.

The difference between the true and the measured background is equal to
the difference between the average of the background and its mode.  The
result of the wrong background determination is equivalent to adding
this difference to the galaxy luminosity measurement.  We find that, on
average, galaxies are measured 0.1 mag brighter in the F814W bandpass
simulations when using isophotal magnitudes.  In the F555W passband,
the shifts depend on the crowding. For NGC 4536, they go from $\sim
0.1$ in the WF4 frame to 0.6 mag for many galaxies fainter than
$m_{F555W} = 22$ in the WF2 frame, the most crowded of the three WFC
frames.  In short, fainter galaxies will be measured up to 0.5 mag too
blue, if using isophotal apertures.

As with the other errors discussed above, this is a systematic error
that will affect both the simulations and the real galaxies, and
therefore will not affect their mean relative reddening. However, for
individual galaxies, the fainter the galaxy the larger the error will
be (in magnitudes). Hence, this effect will increase the error in the
estimate of the mean background galaxy color.  We have reduced this
problem by determining the galaxy colors from a fixed aperture of
0\arcsper5 (5 pixel) diameter.  The use of this aperture size decreased
the shift between the input and the measured color in the simulations
by a tenth of a magnitude, and the dispersion in the shift by up to a
tenth of a magnitude as well.  The error introduced by not measuring
possible color gradients over larger background galaxy radii should be
much smaller than this error, especially when the measurement of a
color gradient depends so critically on an accurate estimation of the
background.  
Much more importantly, by choosing an aperture to perform
the color measurement, we are choosing the scale over which we are
averaging the effects of the obscuring material (\S\ref{discussion}).  At the
distance of NGC 4536, 0\arcsper5 is $\sim 40$ pc; for NGC 3664, the
linear size of the aperture is $\sim 30$ pc. 

\subsection{Systematic errors and the
derived \\extinction law}
\label{syserr}
 
The systematic errors discussed above in ~\S\ref{colbias} and
~\S\ref{bgerr} will affect the simulations and the real galaxies
alike, and hence
will not have an effect on the
reddening determination.  The only error that could have an impact on
the determination of the reddening is the interpolation error in the
HDF $V_{F555W}$ images (\S\ref{HDFdata}).  However, the mean 
expected shift of
$\sim$ -0.03 mag between the interpolated and the 
real $m_{F555W}$ of the HDF galaxies is smaller than 
(1) the dispersion of the shift;
(2) the error in the mean colors of the 
real galaxies and of the sets of output simulated galaxies
(\S\ref{photerr}); and (3) the dispersion in the measurements of
individual galaxy colors(\S\ref{com4536}).  We conclude that the error
introduced by the interpolation will have a negligible effect over both
the background galaxy detection and the reddening determination.

\subsection{The clustering error}
\label{clusterr}

To estimate the contribution from galaxy clustering to the error in the
number-counts, we take the two-point correlation function $w(\theta)$
from Roche et al.\ (1993) or from Brainerd et al.\ (1995), depending on
the completeness magnitude limit of the data.  We assume $(R - I) \sim
0.5$ (Smail et al.\ 1995), since these references list $w(\theta)$ for
limits in the $R$-band.  
Table 1 shows, for each disk
(NGC 4536 and NGC 3664) and field,
the angular size $\theta$ of the field, the assumed (apparent) $R$ magnitude
completeness limit, $w(\theta)$, the number of background galaxies, the
derived $A_I$ (reddened and grey) extinction, the total error in the
extinction, the error contributed by the Poisson error in the
number-counts, and the error owing to clustering errors.  

To estimate how the simulations
contribute to the clustering error,
rather than presenting data for each of the two simulations
used in the calculation of the extinction
(the simulations with numbers of recovered galaxies immediately above
and below the number of real galaxies),
we list data for only one fiducial simulation with the same number
(per WFC field) of recovered HDF galaxies as the real background galaxies. 
Even though the 3 HDF WFC fields have been added
one at a time to each of the WFC fields of the investigated galaxies,
in reality they are of course contiguous, and have to
be treated as one single field for the calculation of the clustering
error. 
The errors contributed by the simulations to the final estimation of
the extinction will be $\sqrt{2}$ the numbers given for the
fiducial simulation.
Except for NGC 4536, WF4, all the other values of
$w(\theta)$ (and therefore of the clustering errors) are slightly
overestimated, both because we take the correlation function for a
square with an area equal to the area under consideration (not a
square), and because Roche et al.\ (1993) include galaxies with
$18 < m_R < 20$ in their calculation of the correlation function; these
galaxies add very significantly to $w(\theta)$, but there are none in
our sample of background objects (we detect no more than a couple in
the simulations without extinction).  Conversely, in the case of the
NGC 4536 WF2 and WF3 fields (analyzed in a combined fashion), there is
a correlated error in the simulations that we have not included, and
that is due to the fact that we count the same HDF fields twice.

\subsection{Photometric error}
\label{photerr}

If the color distribution of galaxies was Gaussian and we were sampling
it in a systematic way, we could confidently equate the error in the
mean color to its measured dispersion divided by $\sqrt{N}$, where
$N$ is the number of galaxies used for the color measurement. However,
neither of these assumptions is true.  The color distribution of
galaxies depends on their limiting magnitude and has an intrinsic
dispersion that is not Gaussian.  Moreover, in our case we have to
infer their mean color from the few galaxies that we can find in the
presence of crowding.  Crowding will have three adverse effects on
the color measurement. The color-magnitude distribution of galaxies
resembles an inverted triangle (cf.\ for example Williams et al.\ 
[1996]). (1) Crowding will push the completeness limit up to brighter
magnitudes, where the intrinsic color dispersion of galaxies is
largest.  (2) It will reduce the number of galaxies available for the
analysis in a dramatic way (Figure \ref{ihist}).  (3) Crowding will
deplete galaxies in an inhomogeneous fashion:  in combination with
clustering, it will determine the random sample of background galaxies
from which we will try to infer the mean color of {\em all} background
galaxies.

To check our estimate of the error in the color measurement, we have
looked at the dispersion in the mean color of 100 random
samples of $N$ galaxies (where $N$ is equal to the number of background
galaxies) taken from the HDF galaxies, to our magnitude limit.  We have
repeated the procedure with the simulations, and we have found that the
predicted error is of the same order as the error found by assuming a
Gaussian distribution.  Most of the time, the error from bootstrap
statistics (Babu \& Feigelson 1996) 
is slightly smaller than the Gaussian error,
probably because the latter is calculated from the real sample and
therefore includes the effect of clustering and other systematics.  In
the case of NGC 4536, we adopt the larger of the two error estimates
for each simulation, as well as for the real galaxies. In the case of
NGC 3664, we use the Gaussian uncertainty.

\section{SUMMARY AND DISCUSSION} \label{summary}

\subsection{NGC 4536} \label{summary4536}

Summarizing the results (errors in parenthesis
are Poisson only, no clustering): the northwestern arm of NGC 4536
shows an attenuation of $A_I = 0.74 \pm 0.49$ 
($\pm$ 0.26) mag (grey extinction) or
$A_I = 1.07 \pm 0.49$ ($\pm$ 0.26) mag (Galactic reddening law),
and a reddening of $E(V_{F555W} - I_{F814W}) < 0.32 $ mag
(grey extinction) or $E(V_{F555W} - I_{F814W}) < 0.29$
mag (Galactic reddening law). The $+1 \sigma$ error of 0.18 mag
is already included in the color uncertainty.
For the northern interarm and outside regions, the results are, respectively,
$A_I <$ 0.46 ($<$ 0.33) mag (grey extinction) or
$A_I <$ 0.52 ($<$ 0.39) 
(Galactic reddening), and 
$E(V_{F555W} - I_{F814W}) = 0.20 \pm 0.13$ mag.
The extinction values for the interarm region already
include a 1$\sigma$ error of $+$0.37 (0.21) mag.
The Poisson errors in parenthesis would be adequate for the
extinction if the
galaxy density was the same behind NGC 4536 as it is in the
HDF.

\subsection{NGC 3664} \label{summary3664}

To summarize the results: the disk of NGC 3664 shows an attenuation 
of $A_I = 1.02 \pm 0.63$ ($\pm$ 0.49) mag (Galactic reddening law) or
$A_I = 1.01 \pm 0.63$ ($\pm$ 0.49) mag (grey extinction),
and a reddening of  
$E(V_{F555W} - I_{F814W}) = 0.40 \pm 0.22$ mag. 
The Poisson errors in parenthesis would be adequate for the
extinction if the 
galaxy density was the same behind NGC 3664 as it is in the
HDF. As it turns out, in the surrounding field of NGC 3664,
this is precisely the case.

\subsection{Conclusions and discussion} \label{discussion}

We have used background galaxy counts and colors, for the first time,
to measure the extinction and reddening through two very different
galaxy disks.  
The images also have very different available exposure times,
which allowed us to test the limits of the method. 

Our principal conclusions are listed below.

(1) In the case of spiral disks, the background galaxy approach is mostly
limited by crowding and confusion in the foreground disk, 
even more than by the background galaxy clustering. We have
developed and calibrated the ``synthetic field'' method to 
decouple the effects of crowding from those of extinction. 
The method comprises the production of artificial frames
by adding suitable and properly scaled reference fields, in this case the HDF,
directly into the spiral galaxy images.
The reference frames are attenuated to mimic different amounts
of extinction and reddening; the idea is to attenuate the control field
until one recovers the same number of simulated galaxies as there are real
background galaxies.
To ensure that systematics and confusion affect synthetic and
real galaxies in the same way, both real and control
galaxies are identified in the same fashion
(i.e., the known positions of HDF galaxies are never used).
The attenuation
derived from the number-counts
should correspond to the average extinction through the
disk of the galaxy.
Finally, a comparison between the average colors of, respectively, 
the control HDF galaxies and the real background galaxies 
will allow an estimate of the degree of reddening.

(2) We tried slightly different lines of attack with both galaxies,
but found them to be fairly equivalent. 
When a $B$ image is
available, like for NGC 3664, its subtraction 
from the $I$ image does seem to facilitate the search for galaxies,
at least in frames with short exposure times. 
It would also be desireable to automate as much as possible
the final identification of real and synthetic galaxies.
Unfortunately,
a fully automatic identification 
might be hard to achieve in frames as crowded as
the ones we are studying. 

(3) There is a significant amount of dust in NGC 3664, even though it
is a Magellanic irregular. 

(4) There is more extinction in the northwestern arm of
NGC 4536 than in the northern interarm region of the galaxy.
The reddenings, however, are comparable.
We have mentioned before that we derive both parameters
independently: the extinction from the galaxy number-counts, and
the reddening from the background galaxy photometry.
The method we use to
measure the extinction will take into account galaxies that are
so attenuated that we cannot see them, while we measure the reddening
from galaxies that are attenuated only to a point where we can
still observe them.
Also, although we average the color excess over a number
of background galaxies, the color of each one of these galaxies is measured
in an aperture with a linear size of $\sim$ 40 pc at the distance
of NGC 4536.
It is possible that at this scale the dust distribution
is genuinely smooth (Keel \& White 1997).
 
In view of all the above, in principle we prefer to present the
results obtained from the number-counts
and from the colors independently, and not 
in the form of a reddening law. The number-counts
measure the total average extinction through a whole region;
the colors measure the extinction owing to a diffuse
component.

On the other hand, if we assume that
nowhere in the region of interest
the extinction is so high that we
cannot sample it with background galaxies,
it is possible to determine an extinction
law from the two values (Appendix). This seems to be a safe
assumption for the northern interarm region of
NGC 4536, and there both the extinction and the reddening
are consistent with a Galactic reddening law.
The similar reddening but significantly higher extinction
in the northwestern arm
can be explained by the combination of a diffuse
obscuring component with regions of higher opacity. 
If, in spite of the caveats, we use the extinction
and the reddening measured
in the northwestern arm to derive a reddening law, 
then the extinction there is greyer than Galactic and
can be caused by unresolved clumps
(Witt \& Gordon 1996), in this case
at the scale of $\sim$ 40 pc.

(5) For the interarm region of NGC 4536, the extinction can be
deprojected to a face-on $A_I <$ 0.22 mag.
For the northerm arm of NGC 4536, however, 
given the patchy dust distribution,
it is not possible to
derive a number for the face-on extinction in a model-independent
way. 
We can, however, assume that the 
reddening of the background galaxies behind the arm
is caused by a smooth component and follows
a Galactic extinction law.  In that case, the reddening
implies a face-on extinction through the diffuse component 
in the arm of
$A_I <$ 0.16 mag.  

(6) The usefulness of the background galaxy counts and colors 
to measure foreground extinction has been recently confirmed by
their application to the field of the gamma-ray burster 
GRB970228 (Fruchter et al.\ 1998).
In that analysis, we used four different reference frames
to measure the extinction {\em in the F606W passband}: 
the HDF; the weak radio galaxy 53W002 field;
and two fields located, respectively, 
at J(2000) 15:58:49.8, 42:05:23, and at 
J(2000) 14:17:43.63, 52:28:41.2 
(Westphal, Kristian, \& Groth 1994). 
Since the GRB970228 field is not crowded, it was not necessary
to produce ``synthetic fields". Under those circumstances, 
and assuming a Galactic reddening law, 
both counts and colors yield the same result
(respectively, 0.42$\pm$0.23 mag, and 0.54$\pm$0.04 mag). 
The average of these two measurements 
of the Galactic extinction towards the gamma-ray
burster 
(0.48$\pm$0.12 mag) is consistent with the value of\ \ 0.72\\
$\pm$0.24 mag obtained by
Burstein \& Heiles (1982), and   
with the extinction of 0.69$\pm$0.11 measured from 
maps of dust IR emission (Schlegel, Finkbeiner,
\& Davis 1997). The marginal systematic difference
in the measurements can be explained by our better
spatial resolution (roughly 1$^\prime$, vs. 
6\arcmper1 of Schlegel et al.\ and 0\degper6 of
Burstein \& Heiles), since the burster is located at a
position ($l$ = 188.913\deg, $b= -17.941\deg$)
where the Milky Way exhibits a steep gradient in extinction
(cf.\ Burstein \& Heiles 1982). 

(7) From the Cepheids detected in the same NGC 4536 data, 
Saha et al.\ (1994) derive a mean extinction 
of 0.07$\pm$0.05 mag 
in Kron-Cousins $I$\footnote{From 
Figure 12 of Holtzman et al.\ (1995), 
the extinction in Kron-Cousins $I$ is directly comparable with
the extinction in $I_{F814W}$.}. 
Likewise, 
they estimate the $I$ differential extinction 
between the arm and the interarm regions to be 
0.05$\pm$0.10 mag. 
These numbers imply for the
arm of NGC 4536 an extinction $A_I$\lax 0.2 mag, 
much lower than what we find from the galaxy number-counts.
More importantly, the detected Cepheids span
a range of at least 2 magnitudes in brightness;
Saha and co-workers had the capability to measure
an extinction much higher than 0.2 mag. The fact that they did not
implies that the Cepheids detected in the arm lie preferentially either 
in areas of low extinction or on top of the
obscuring material. 
Together with the extinction derived from the
galaxy number-counts, the Cepheid reddening  
is consistent with the existence of  
regions of significant opacity in the arm of NGC 4536.
On the other hand, the extinction measured
from the Cepheids agrees 
with the one we find for the interarm region and, from the
galaxy colors, for 
the diffuse component in the arm
($<$0.52 and $<$0.42 mag, respectively, for a Galactic reddening model).
Lastly, since the Cepheids are embedded in the disk,
statistically they should suffer only about half of the total extinction,
which background illuminating objects can probe;
our measurements and
those by Saha and collaborators are consistent with this picture.

(8) That in the case of spiral galaxy disks we are totally dominated
by crowding is demonstrated by the fact that there is 
less than 30 percent improvement in
the extinction errors for NGC 4536, relative to 
NGC 3664, even though we have 5 times better 
signal-to-noise ratio in $I_{F814W}$  
and more than 10 times better signal-to-noise ratio in 
$V_{F555W}$. Likewise, the color errors are only 20 percent better
for NGC 4536.  Perhaps even more telling, 
the accuracy of both the extinction and
the color measurements is the same in the less obscured regions of
the galaxies, despite the enormous difference in exposure times. 

We can both reduce the contribution to the 
relative errors from the control fields,
and perhaps even measure the clustering error directly,
by using an increasing number of them. 
And since the dominating source
of error is crowding, it certainly will not matter if we use 
reference fields with shorter exposure times than the HDF to produce
the ``synthetic fields''.  However, even if we manage to 
determine very accurately the mean and the dispersion of the number of 
galaxies in a typical WFPC2 reference field,
we will still be left with 
the larger source of error:
the studied galaxies themselves, 
with their ``single realization'' background fields.  
If NGC 4536 and
NGC 3664 are representative, the extinction errors  
will go from 0.3 mag in galaxies with exposures 
of several $\times$ 10$^4$ seconds, to 0.5 mag in galaxies 
with shallow exposures of a few $\times$ 10$^3$ seconds;
the error in the color will amount to $\sim$ 0.1 mag in all cases.
In the future we will try to reduce this source of error by 
combining together the results from background galaxies behind
spirals of similar Hubble type, inclination, and distance
to the Milky Way. 

\acknowledgements
R.A. Gonz\'alez is grateful to Jay Anderson for his cosmic-ray removal
program, to Abi Saha for his advice on how to deal with the images of
NGC 4536, and to Stefano Casertano and Andy Fruchter for useful
discussions on galaxy clustering and bootstrap statistics.  Larry Petro
kindly confirmed that the images of NGC 4536 were not affected by
scattered light.  
We want to thank M. Dahlem and S. M. Baggett for providing the image of
NGC 3664.  B. Dirsch is especially grateful to  S. M. Baggett for
her help in the reduction of the WFPC2 images of NGC 3664.
We thank the referee, W.E. Keel, whose comments helped us
improve the presentation of the manuscript.
Support for this work was provided partly by NASA
through grant AR-06400.01-95A and partly by the Director's
Discretionary Research Fund at the Space Telescope Science Institute.

\newpage   
\onecolumn

\begin{appendix}

\centerline{APPENDIX}

\centerline{METHOD-DEPENDENT GREYNESS?}

The reddening and the attenuation are derived independently with our approach. 
In what follows, we discuss the limitations of the method
for the derivation of a meaningful extinction law.

Let us consider the case where each of the background galaxies used to
derive the ``average'' extinction through the disk
suffers from a different amount of Galactic reddening. 
Our method will derive a reddening law that can be
described by the following equation:

\begin{equation}
\overline{R_I} = \frac{\overline{A_I}}{\overline{E(V-I)}},
\end{equation}

\noindent where $\overline{A_I}$ denotes the extinction
derived from the number-counts and 
$\overline{E(V-I)}$ is the
average reddening of the recovered galaxies. 
$\overline{E(V-I)}$ can also be written as:

\begin{equation}
\overline{E(V-I)} = \frac{1}{N}\sum_{i} {E(V-I)}_i = 
\frac{1}{N R_I}\sum_{i} {A_I}_i,
\end{equation}

\noindent where the summation encompasses all the background galaxies.

The extinction, on the other hand, can be expressed as (\S\ref{results}):

\begin{equation}
A_I = -2.5 \cdot C \cdot log(\frac{N}{N_0}),
\end{equation}

\noindent where $C$ describes selection biases and $N_0$ is the 
normalization.

If $P(m_i,{A_I}_i)$ is the probability (0 or 1)  
of finding a galaxy with magnitude $m_i$ 
and suffering an attenuation ${A_I}_i$, 
we can rewrite the above equation as:

\begin{equation}
\overline{A_I} = -2.5 \cdot C \cdot log(\frac{1}{N_0}\sum_{i}P(m_i,{A_I}_i))
\end{equation}

To obtain the attenuation in a fractional area  
$f_j$, with attenuation ${A_I}_j$, of the original image, the equation becomes:

\begin{equation}
{A_I}_j = -2.5 \cdot C \cdot log(\frac{1}{f_j N_0}\sum_{i}P(m_i,{A_I}_j)).
\end{equation}

Splitting the summation in equation (4) into two summations, one running over 
the areas with a given attenuation $A_j$,
the other running over all galaxies in each of these areas, 
equation (4) becomes:

\begin{equation}
\overline{A_I} = -2.5 \cdot C \cdot log(\frac{1}{N_0}\sum_{area,j}\sum_{i}P(m_i,{A_I}_j)).
\end{equation}

Using the value of $\sum_{i}P(m_i,{A_I}_j)$ from 
equation (5), and assuming that 
$\sum_{area,j}f_j = 1$, we finally obtain: 

\begin{equation}
\overline{A_I} = -2.5 \cdot C \cdot log(\sum_{area,j}f_j 10^{\frac{-0.4}{C}{A_I}_j}).
\end{equation}

The ``reddening law'' would read:

\begin{equation}
\overline{R_I} = \frac{-2.5 \cdot C \cdot log(\sum_{area,j}f_j 10^{\frac{-0.4}{C}{A_I}_j})}
{\frac{1}{N}\sum_{i}{A_I}_i}R_I
\end{equation}

Equation (7) shows that the extinction obtained from the number-counts
is a linear average 
{\em over the whole area considered\footnote{That is, the whole
area that is sampled by the reference galaxies, so 
crowding will not affect the extinction measurement}, and as such
takes into account real background galaxies that we cannot see}, 
while the color excess 
involves the average of the absorption in magnitudes,
{\em and only over background galaxies that we 
actually see}. 
Strictly, the measured $\overline{R_I}$ and the intrinsic
$R_I$ will be equal only if $A_I$ is the same for all galaxies.   
But, mainly, 
it is possible to measure a much higher extinction
from the number-counts than the extinction implied by the
reddening.
\end{appendix}
\newpage
\twocolumn

\vfill
\eject

\newpage
\onecolumn
\tighten

\vfill

\begin{table}[tp]
\vspace{-1.8in}
\hspace*{-1.2in}\psfig{figure=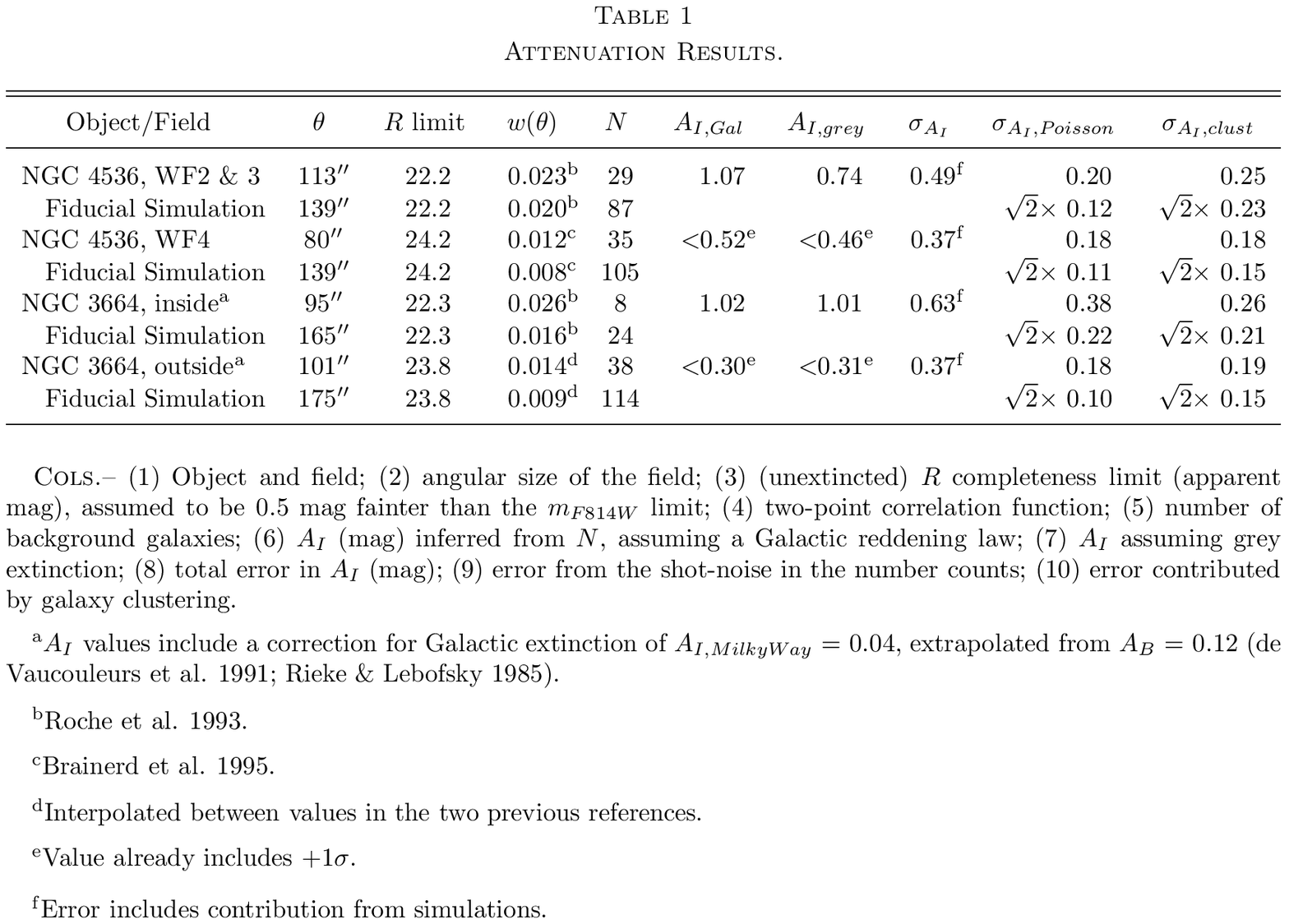,width=9in,angle=0}
\label{tclusterr}
\end{table}

\newpage
\begin{table}[tp]
\vspace{-2.2in}
\hspace*{-1.2in}\psfig{figure=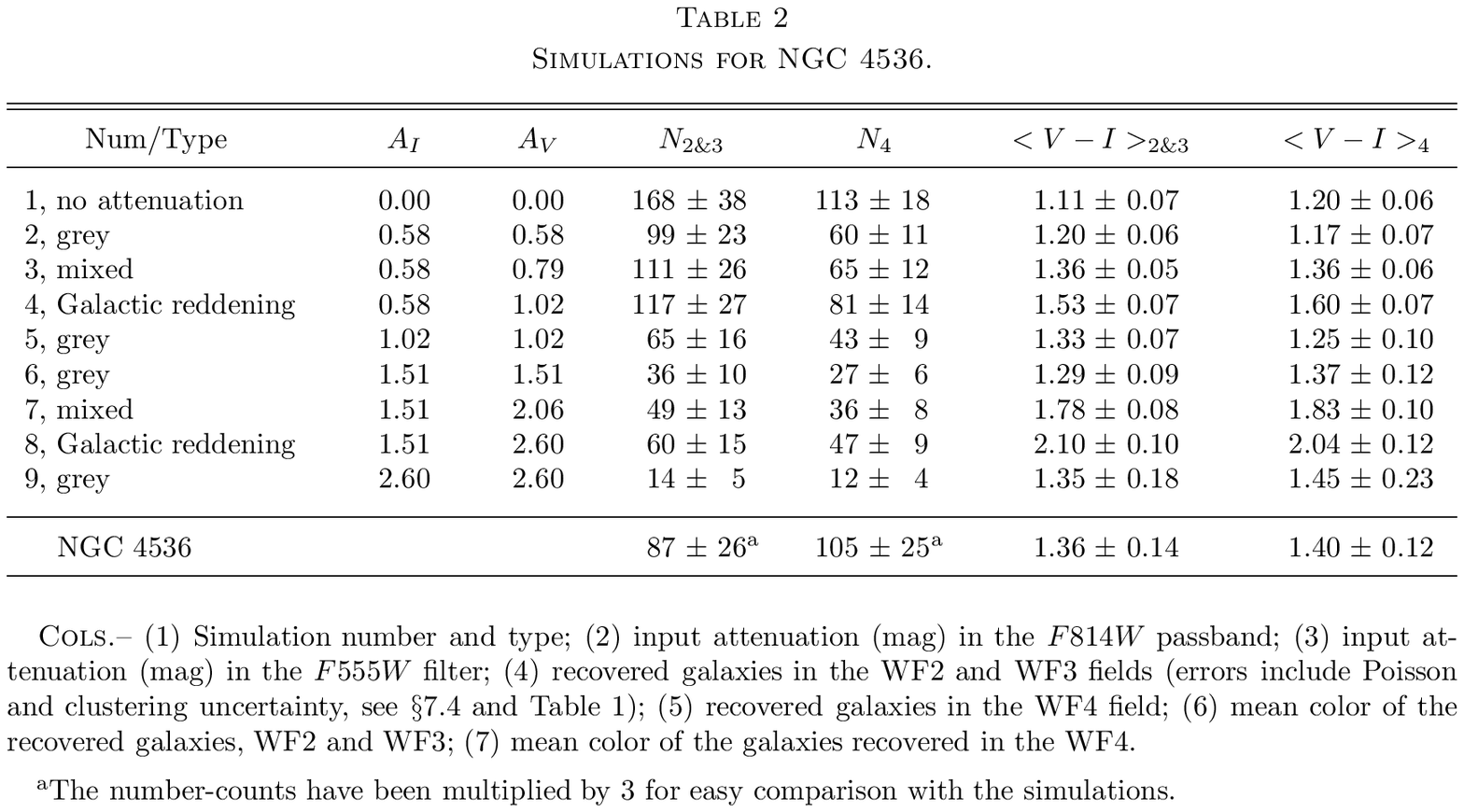,width=9in,angle=0}
\label{tsim4536}
\end{table}

\newpage
\begin{table}[tp] 
\vspace{-2in}
\hspace*{-1.2in}\psfig{figure=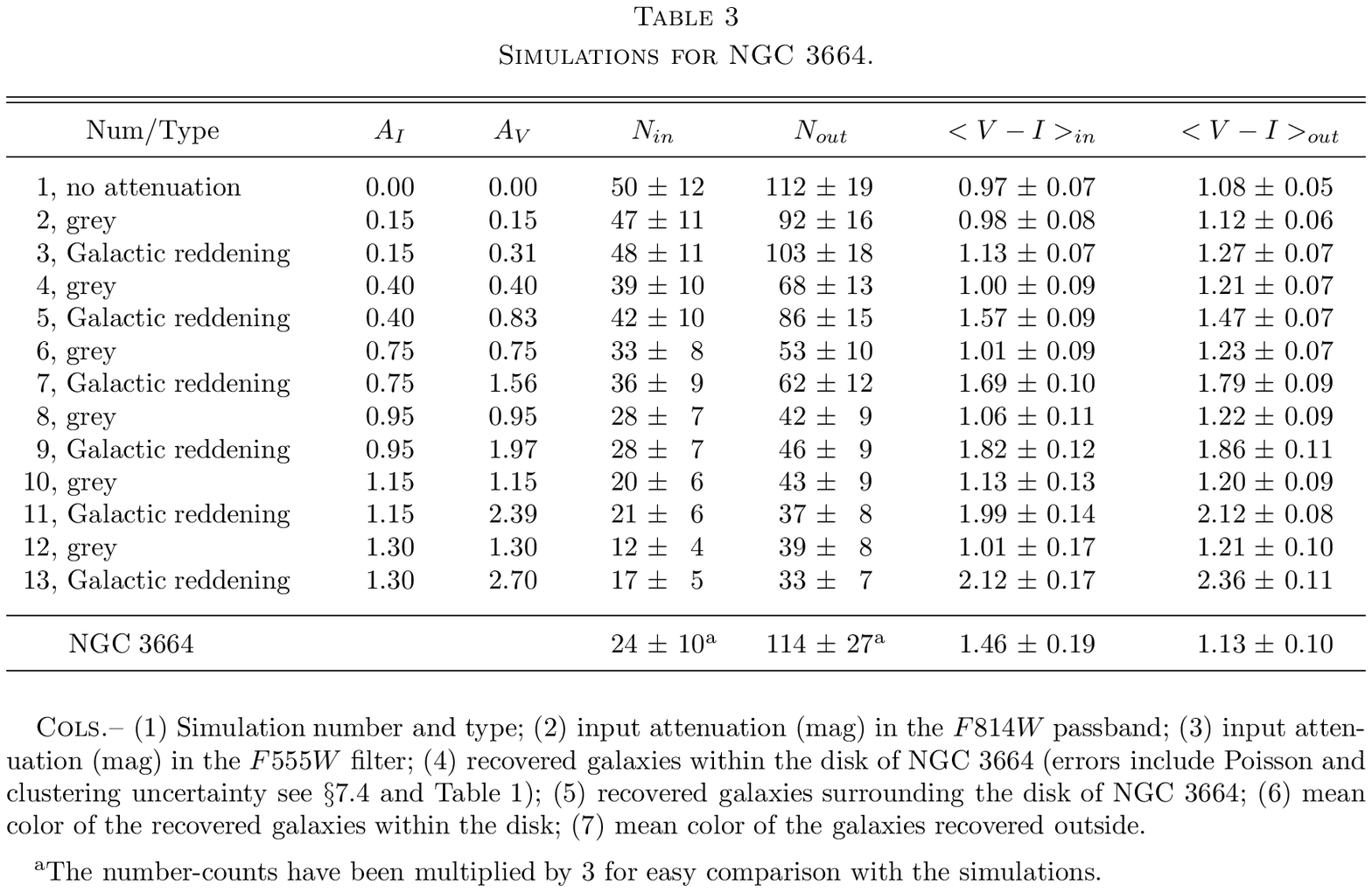,width=9in,angle=0}
\label{tsim3664}
\end{table}

\clearpage

\newpage

\begin{figure}
\plotfiddle{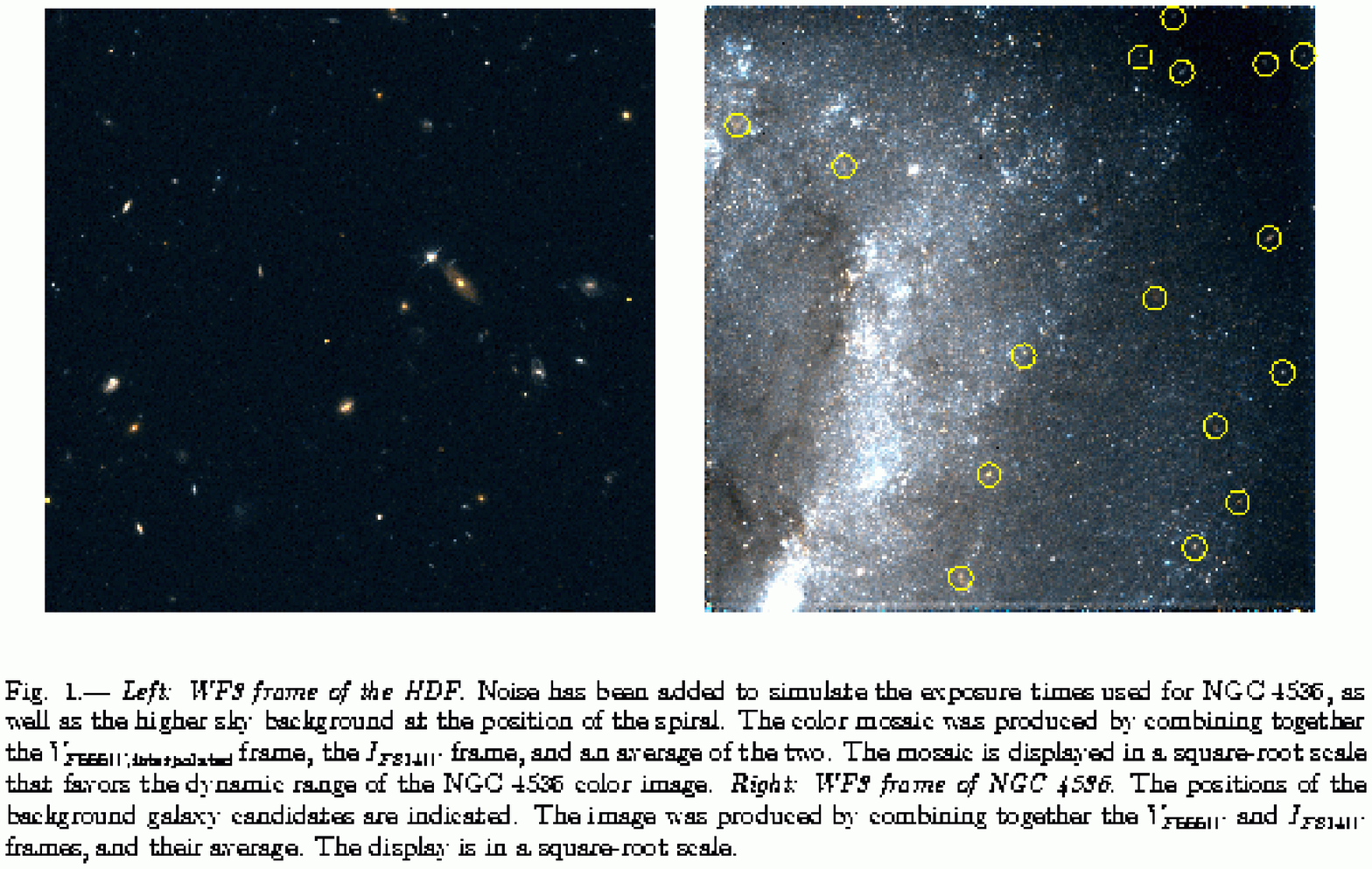}{7.3in}{90}{100}{100}{390}{-20}
\label{colorpic}
\end{figure}

\begin{figure}
\hskip 0.5in \psfig{figure=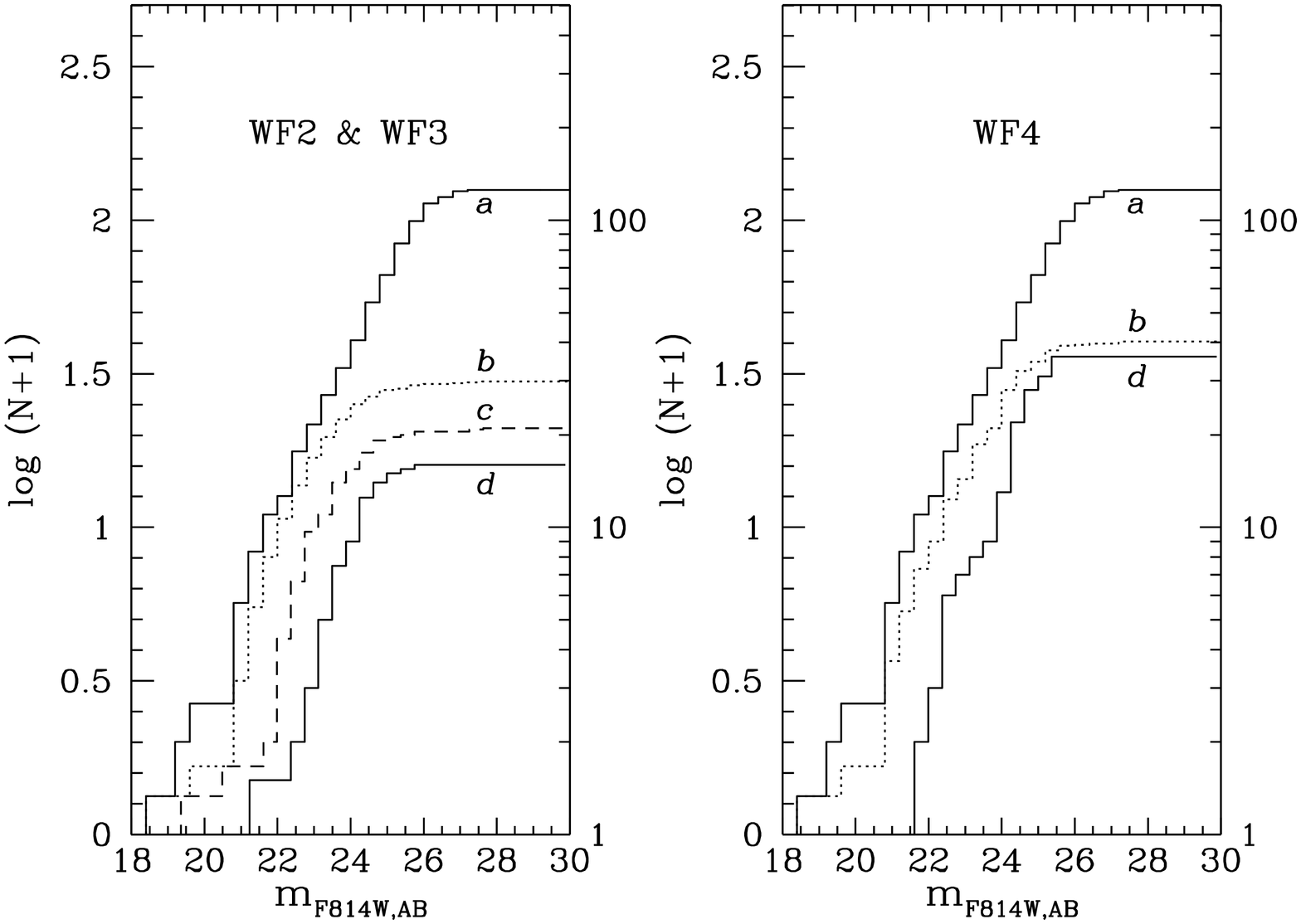,width=6.5in,angle=-90}
\caption{Cumulative histograms of galaxy counts
in $I_{F814W}$ images. {\em Left:} WF2 and WF3. {\em A:} Field galaxies
in the ``degraded'' HDF; {\em b:} HDF galaxies recovered from the
synthetic fields without extinction; {\em c:} HDF galaxies recovered
from the simulation with 0.6 mag of extinction
(Galactic reddening curve) at $I_{F814W}$;
{\em d:} Real background galaxies in the NGC 4536 frames. The numbers
of galaxies have been normalized to the area of one WFC field.
{\em Right:} WF4. Histograms labeled as in left panel (no simulation
with extinction is included).}
\label{histograms}
\end{figure}

\begin{figure}
\hskip 0.5in \psfig{figure=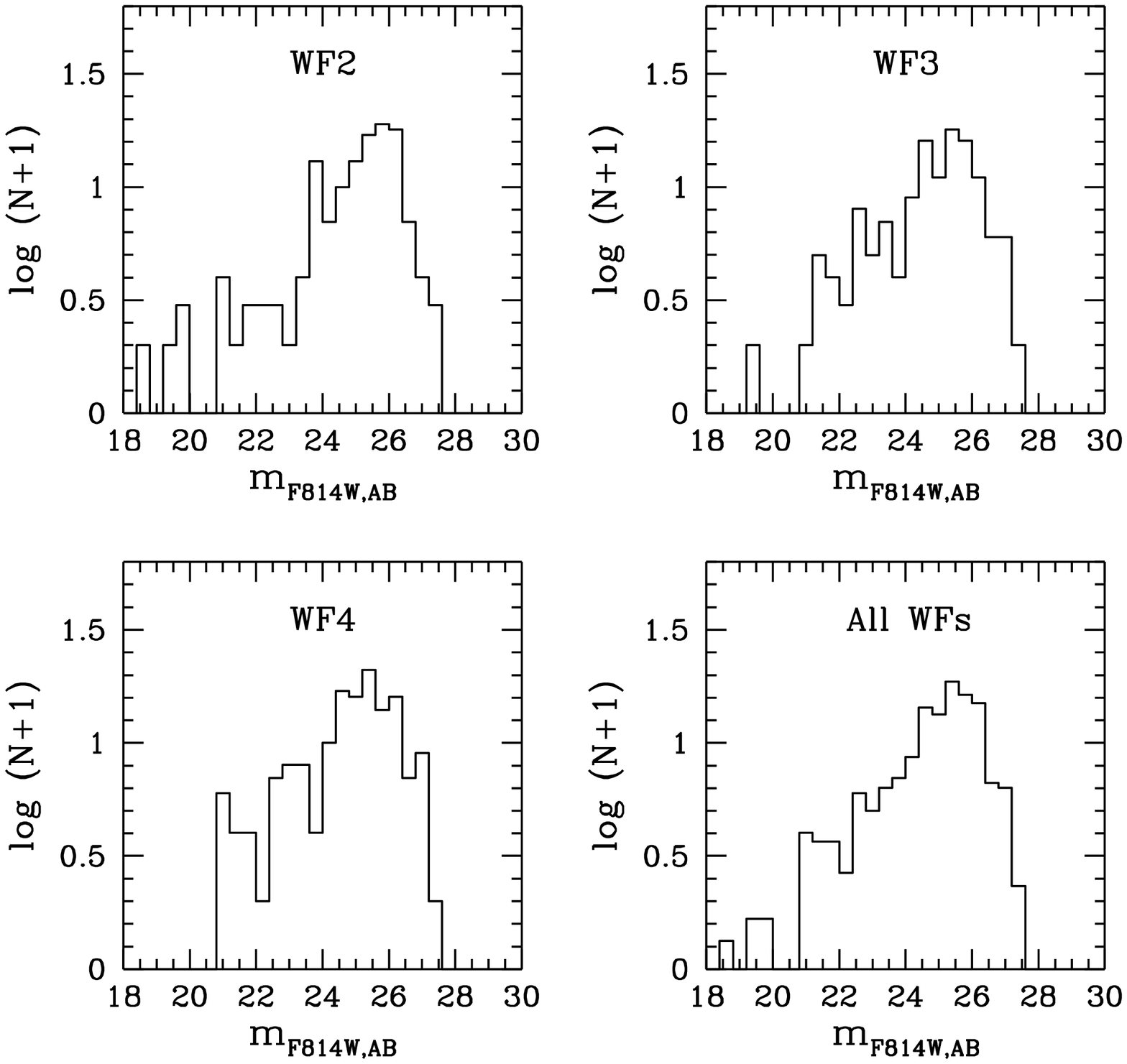,width=6.5in,angle=0}
\caption{Number of HDF galaxies per (isophotal) $m_{F814W}$ bin. 
The HDF's signal-to-noise ratio has been
degraded to the level adequate to the exposure time and sky background of
the NGC 4536 image. {\em Top left:} WF2; {\em top right:} WF3;
{\em bottom left:} WF4; {\em bottom right:} average of the 3 WF fields.
In general, comparisons of real and simulated background galaxies will be done
against this average.}
\label{fhdfi}
\end{figure}

\begin{figure}
\hskip 0.5in \psfig{figure=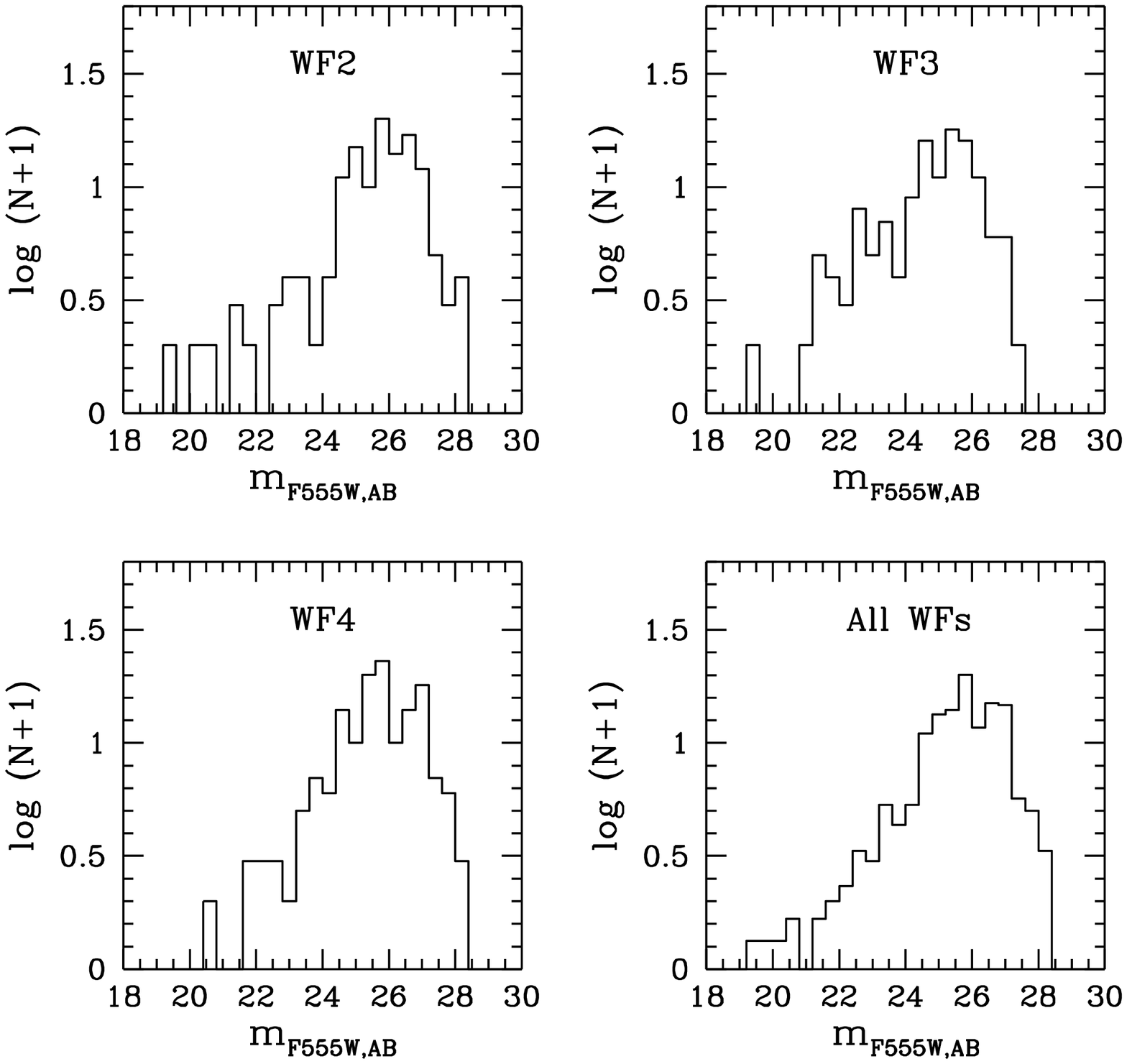,width=6.5in,angle=0}
\caption{Same as figure \ref{fhdfi}, for the $V_{F555W}$ filter.
The galaxies were searched for in the $I_{F814W}$ images, so 
they are the same objects shown in figure \ref{fhdfi},
and their light has been integrated up to the same limit.}
\label{fhdfv}
\end{figure}

\begin{figure}
\plotone{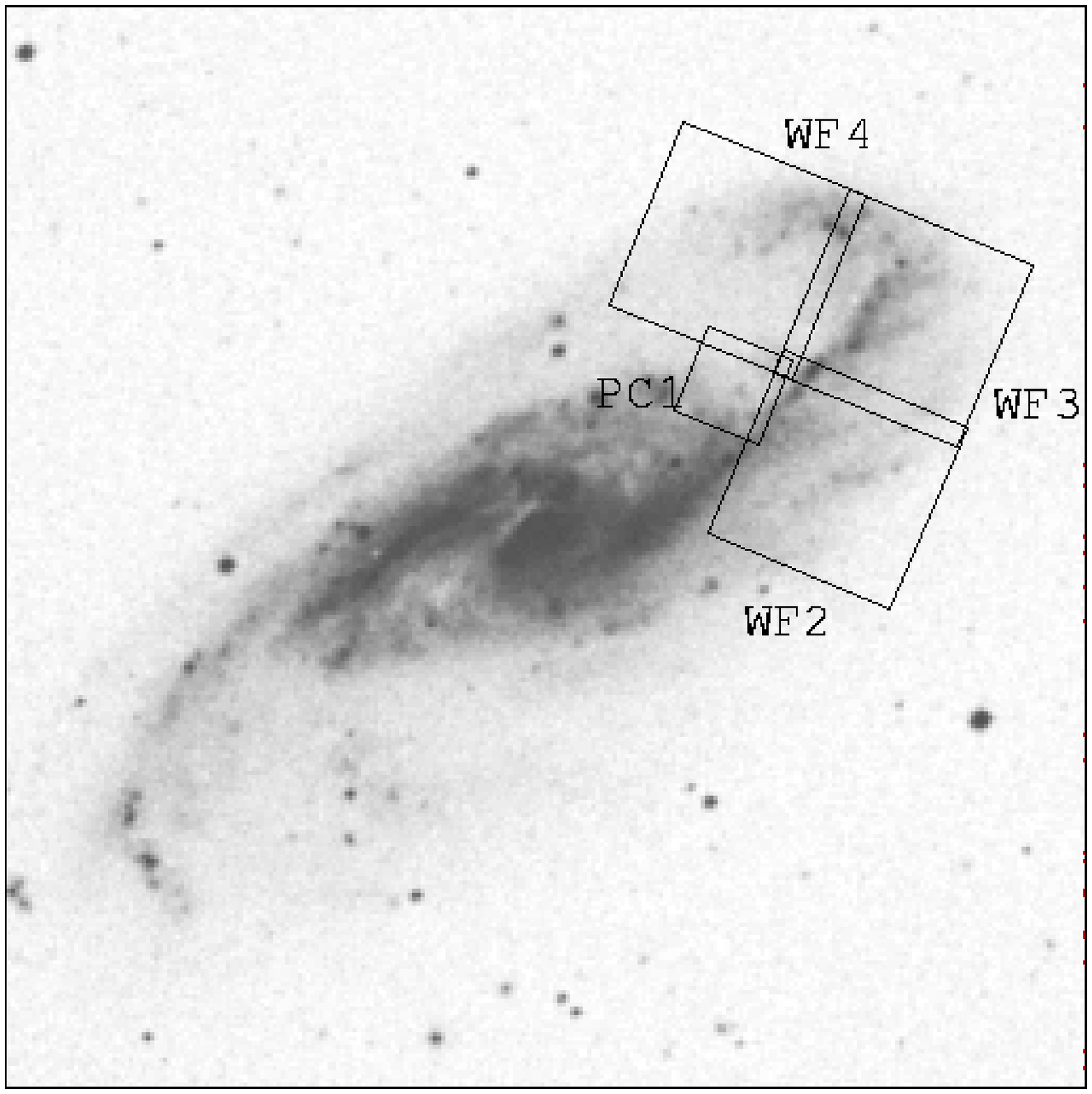}
\vspace*{0.2in}
\caption{STScI Digitized Sky Survey image of NGC 4536,
displayed with a linear scale. 
The blue (J) Science and Engineering Research Council
Survey plate was obtained on 1979 May 19, 
with the UK Schmidt Telescope.
The position of the 
WFPC2 field is superimposed. 
North is up and East is to the left.
}
\label{fn4536}
\end{figure}

\begin{figure}
\hskip 0.5in \psfig{figure=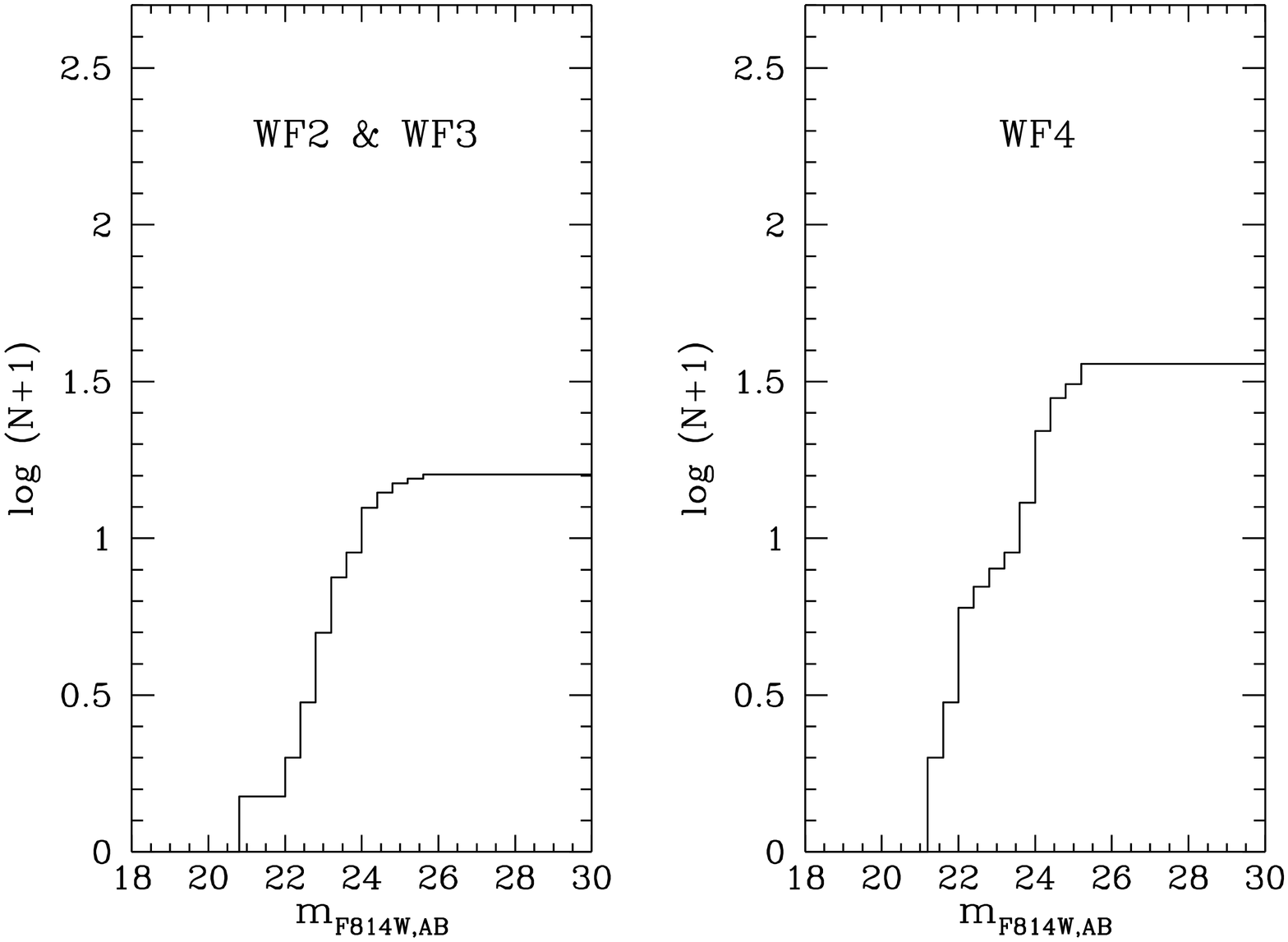,width=6.5in,angle=-90}
\caption{Cumulative histograms of real background galaxies 
found in the NGC 4536 $I$ images. {\em Left:} WF2 and WF3. 
The numbers have been
normalized to the area of one WFC field. {\em Right:} WF4.} 
\label{cbgi4536}
\end{figure}

\begin{figure}
\hskip 0.5in \psfig{figure=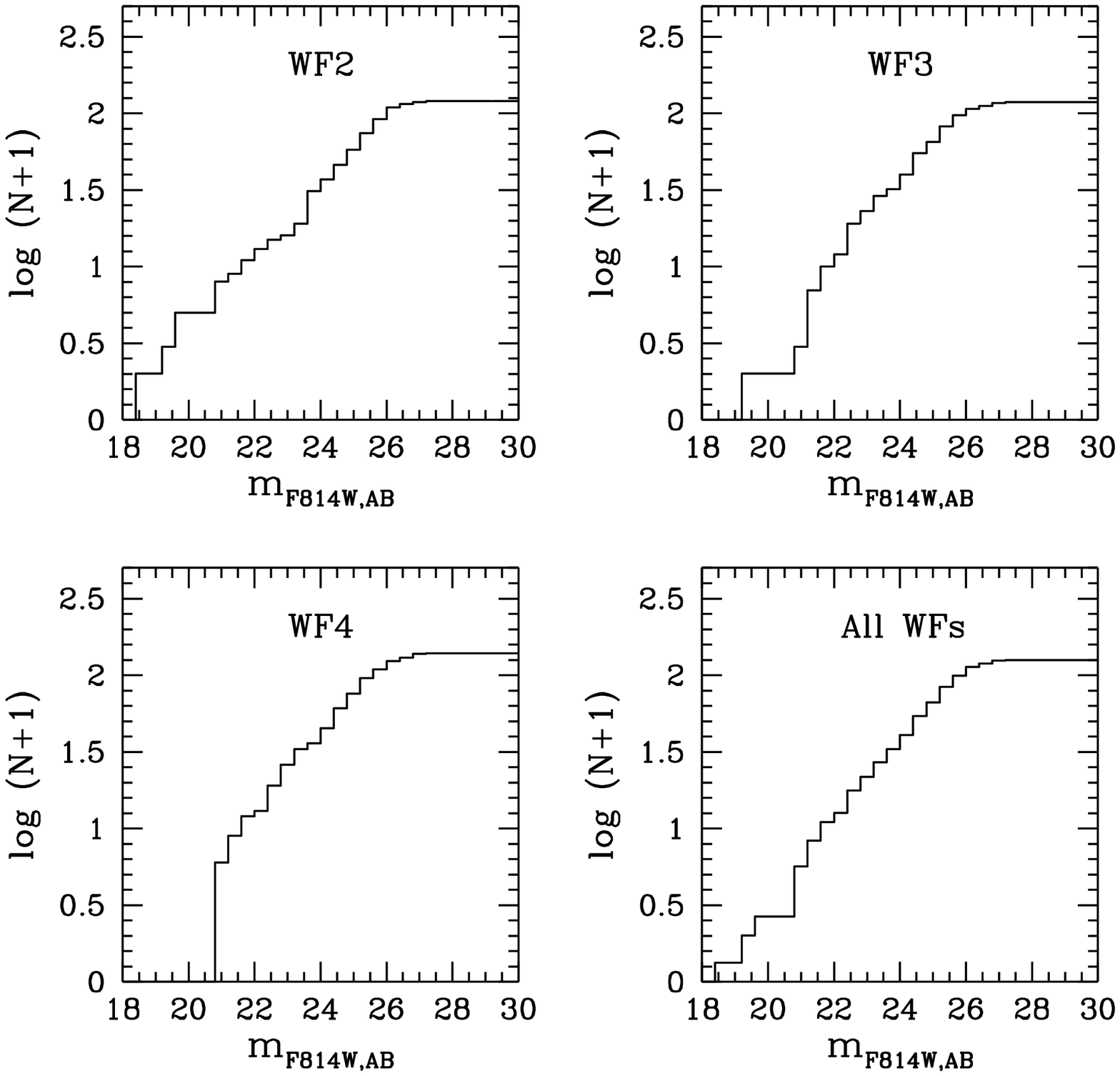,width=6.5in,angle=0}
\caption{Cumulative histogram of the galaxies in the ``degraded'' HDF
$I_{F814W}$ images. {Top left:} WF2; {\em top right:} WF3;
{\em bottom left:} WF4; {\em bottom right:} average of the 3 WF fields.
The differences in the number-counts owing to galaxy clustering
are most noticeable at (isophotal) magnitudes brighter than
$m_{F814W} \sim$ 24.} 
\label{cumi}
\end{figure}

\begin{figure}
\hskip 0.5in \psfig{figure=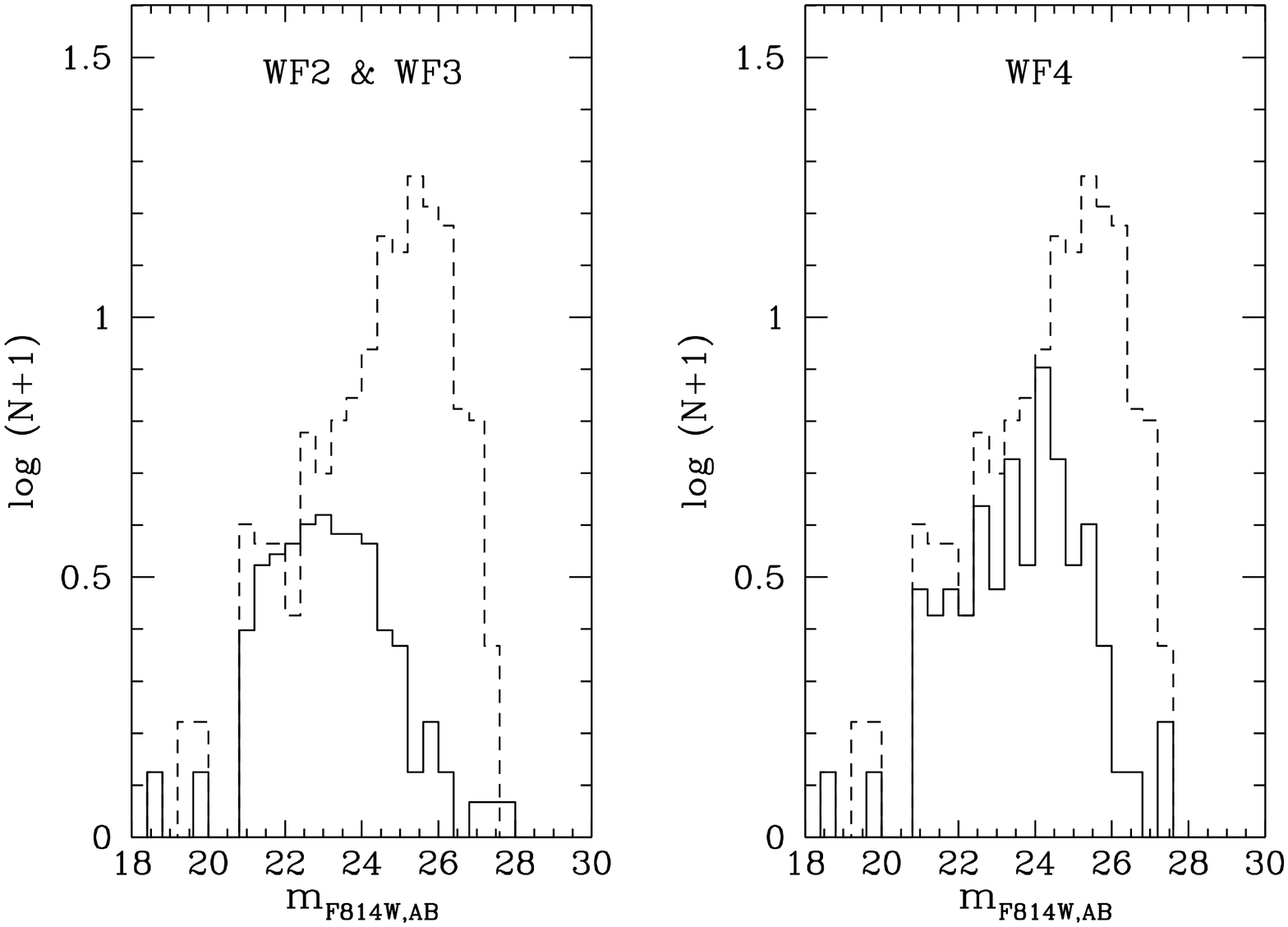,width=6.5in,angle=-90}
\caption{Comparison between the average number of HDF galaxies recovered
from the NGC 4536 frames in the synthetic field without extinction
({\em solid line}), 
and the average number per (isophotal) $m_{F814W}$ 
bin of all HDF galaxies ({\em dashed line}). 
The HDF's signal-to-noise ratio has been
degraded to the level adequate to the exposure time and sky background of
the NGC 4536 image. The figure shows the dramatic effect of crowding
alone on the number counts. {\em Left: WF2 and WF3.} 
Given the isophotal brightness limit, 
while the HDF galaxies are complete to at
least $m_{F814W}$ = 25, when added to the NGC 4536 they are
complete only to $m_{F814W}$ = 23. The numbers have been
normalized to the area of only one WF field. {\em Right: WF4.} 
Simulation is complete to $m_{F814W}$ = 24.
}
\label{ihist}
\end{figure}

\begin {figure}
\hskip 0.5in \psfig{figure=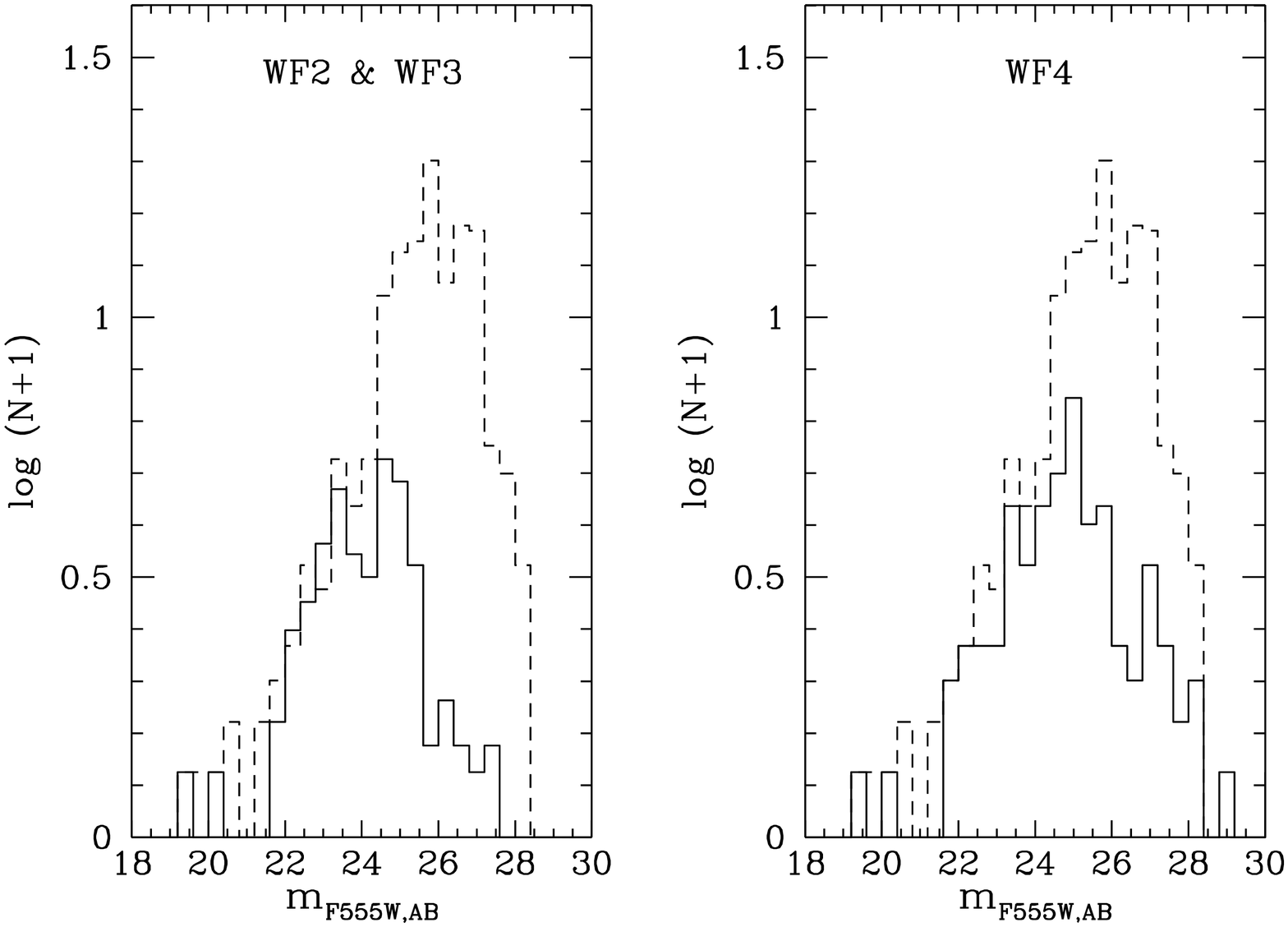,width=6.5in,angle=-90}
\caption{Same as figure \ref{ihist}, for the $V_{F555W}$ passband.
The objects in the synthetic fields were searched 
for in the $I_{F814}$ frames, so they are
the same as those shown in figure \ref{ihist}.
Many more faint, blue galaxies were detected in the WF4 NGC 4536
frame ({\em right}), relative to the WF2 and WF3 fields
combined ({\em left}).
}
\label{vhist}
\end{figure}

\clearpage

\begin {figure}
\hskip 0.5in \psfig{figure=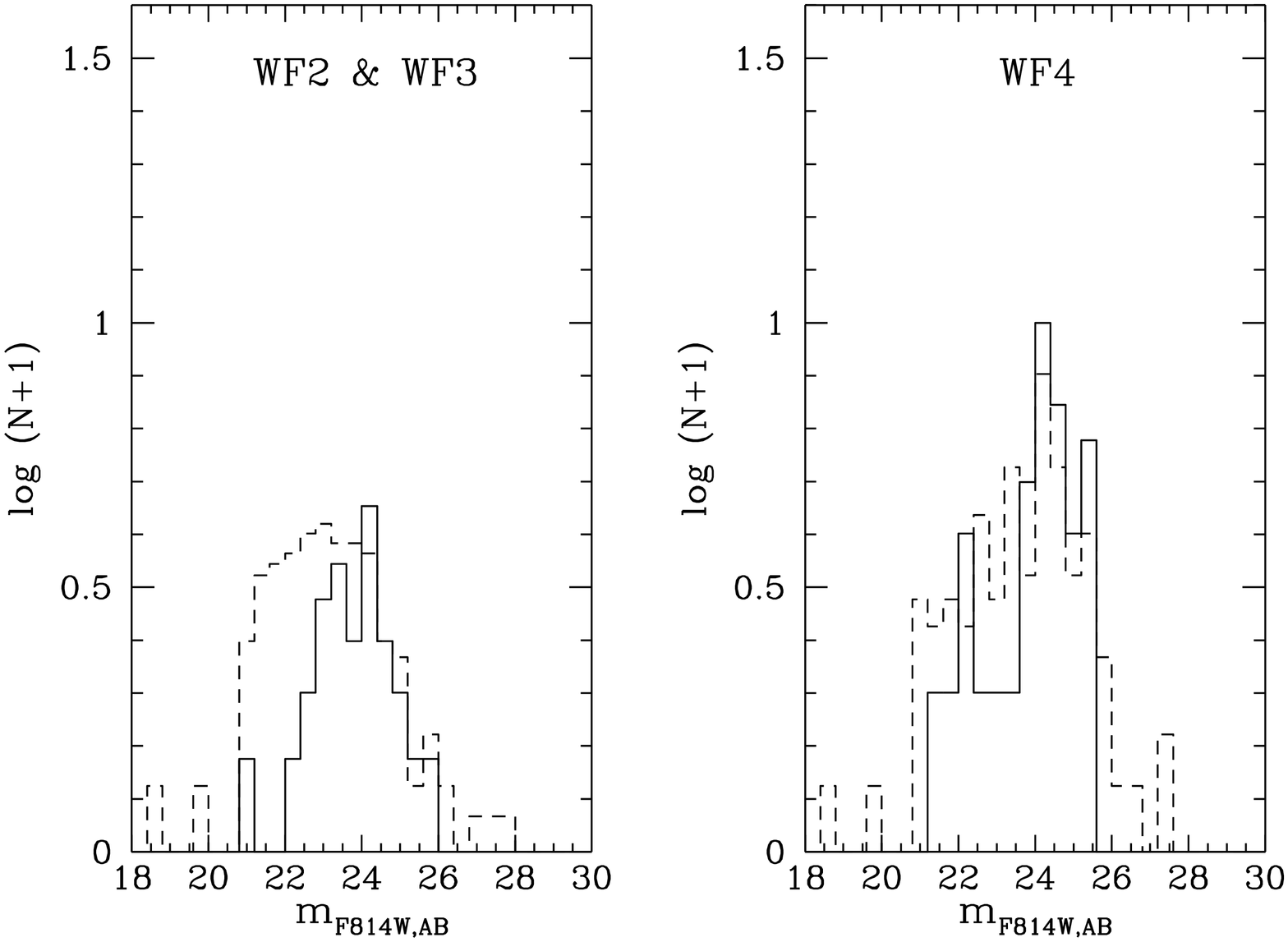,width=6.5in,angle=-90}
\caption{Comparison between the number of background galaxies 
({\em solid line}) and the average number per (isophotal) 
$m_{F814W}$ bin of HDF galaxies recovered
from the NGC 4536 frames in the synthetic field without attenuation
({\em dashed line}). {\em Left: WF2 and WF3.} 
The effects of extinction are readily noticeable,
both from the reduction in
the number-counts and from the shift to fainter magnitudes
of the real galaxies. The numbers have been normalized to
the area of only one WF field. {\em Right: WF4.} Any effects 
are much less apparent.}
\label{simreali}
\end{figure}

\begin {figure}
\hskip 0.5in \psfig{figure=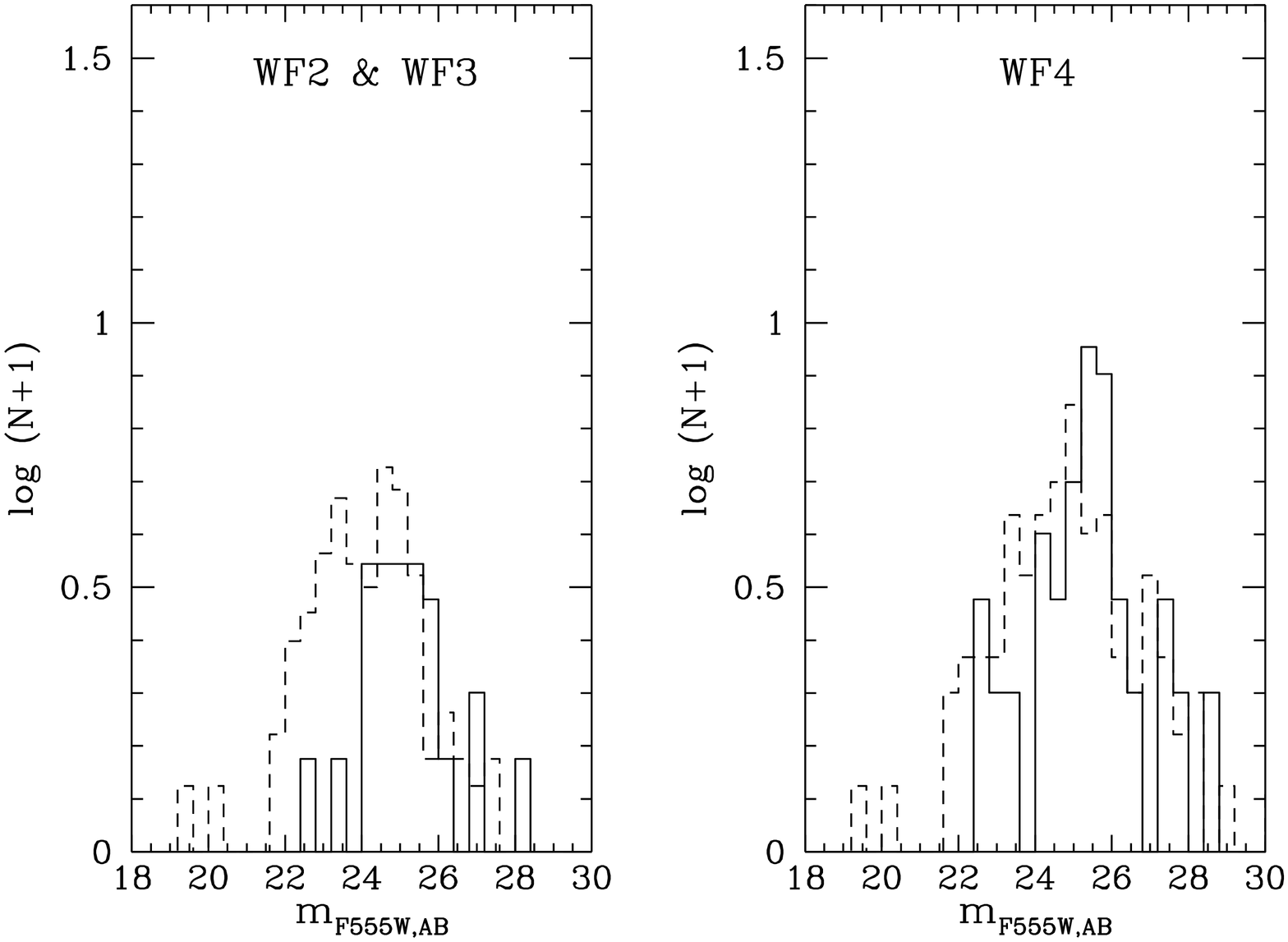,width=6.5in,angle=-90}
\caption{Comparison between the number of background galaxies
({\em solid line}) and the average number per 
$m_{F555W}$ bin of HDF galaxies recovered
from the NGC 4536 frames in the synthetic field without attenuation
({\em dashed line}). The galaxies were searched for in the
$I_{F814W}$ images, and hence they are the same as those
in figure \ref{simreali}. {\em Left: WF2 and WF3.}
Both the numbers and the brightnesses of the background galaxies 
are reduced, compared to the simulation. The numbers have been
normalized to the area of only one WF field. {\em Right: WF4.} Effects
of extinction, especially in the numbers, are much less apparent.}
\label{simrealv}
\end{figure}
 
\begin {figure}
\hskip 0.5in \psfig{figure=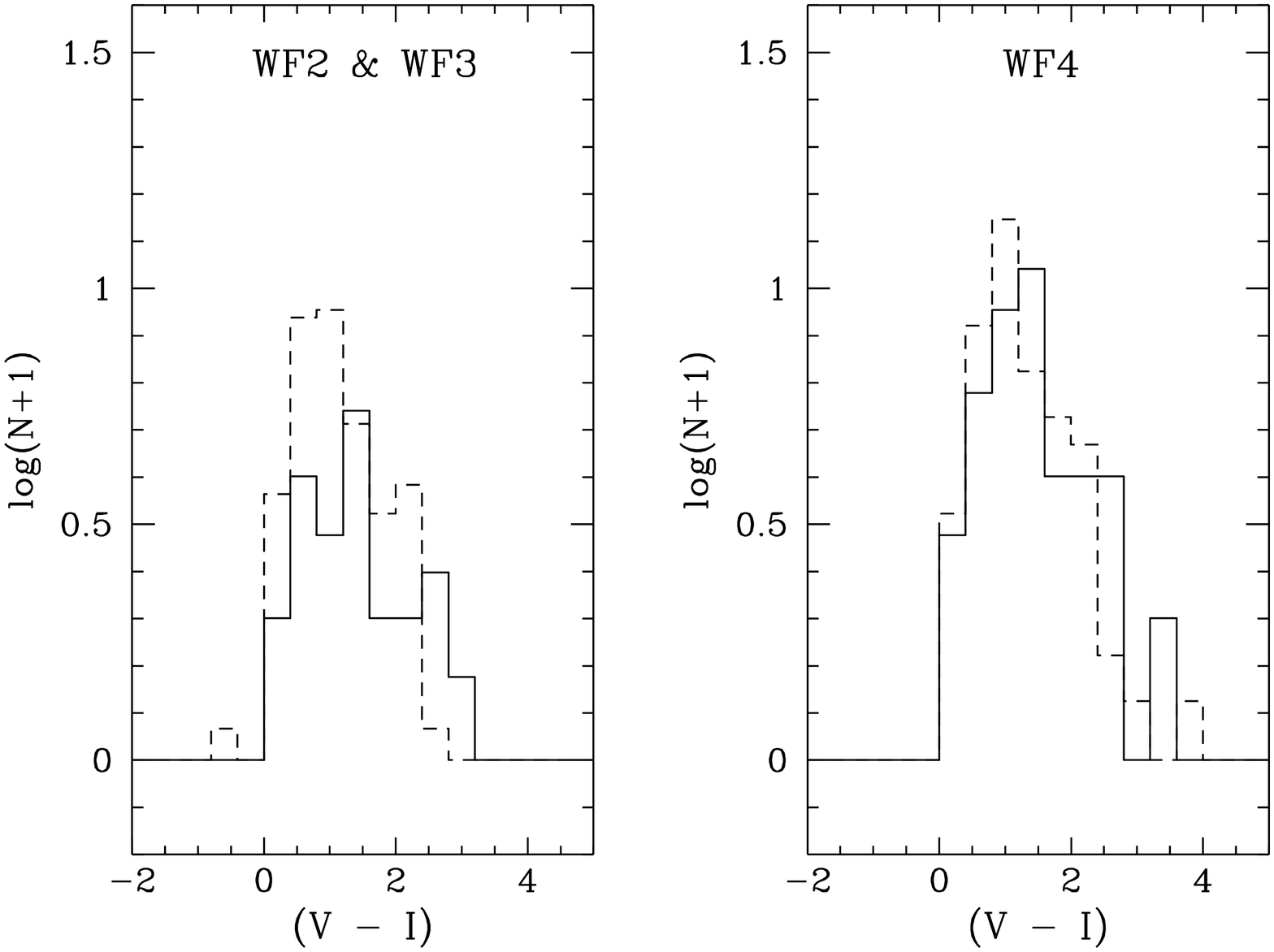,width=6.5in,angle=-90}
\caption{Comparison between the ($V_{F555W} - I_{F814W}$) color 
of the background galaxies ({\em solid line}) 
and the color of the HDF galaxies recovered
from the NGC 4536 frames in the synthetic field without attenuation
({\em dashed line}). {\em Left: WF2 and WF3.} Their numbers look depleted,
but the effects of extinction on the color of the real galaxies are not 
very noticeable. The numbers have been normalized to the area of 
only one WF field. {\em Right: WF4.} Both the numbers and the colors of
the real galaxies are consistent with those of the simulation. 
}
\label{simrealcol}
\end{figure}

\begin {figure}
\plotfiddle{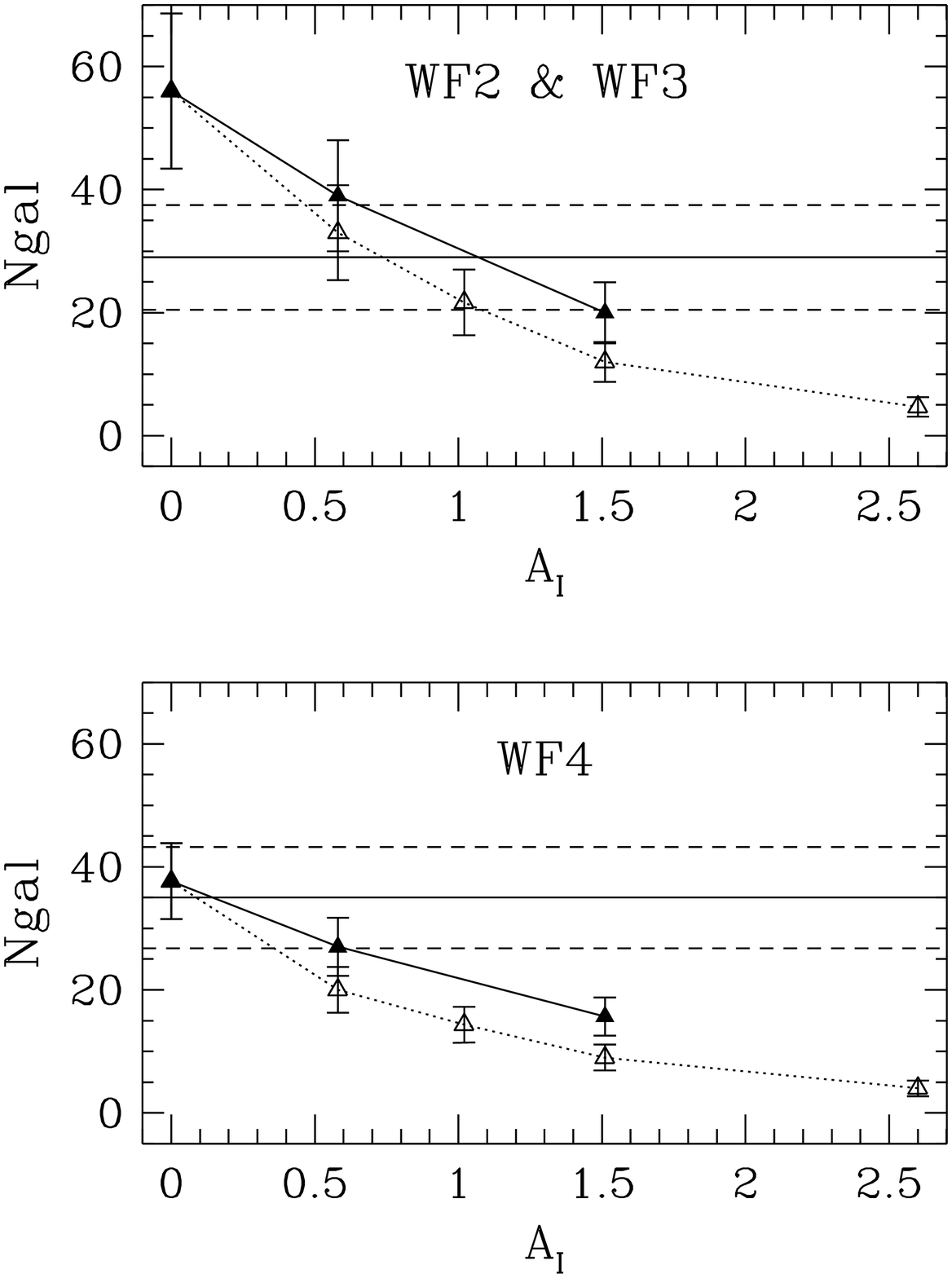}{6.3in}{0}{60}{60}{-210}{0}
\caption{{\em Top: NGC 4536, WF2 and WF3 frames.
Filled triangles and solid line:} average number of
recovered HDF galaxies,
vs. simulated Galactic reddening extinction of $A_I$.
{\em Open triangles and dotted line:} average number of recovered
HDF galaxies vs. extinction, when using a
grey extinction model for the simulations.
The difference between the extinction inferred from the
two models increases with extinction.
The error bars include the statistical Poisson uncertainties,
as well as the field-to-field variations expected from
galaxy clustering. 
{\em Horizontal solid line:} number of real background galaxies
with its uncertainty ({\em horizontal heavy dashed lines}).
{\em Bottom: NGC 4536, WF4 frame.} Symbols as in top panel.}
\label{fext4536}
\end{figure}

\begin {figure}
\hskip 0.5in \psfig{figure=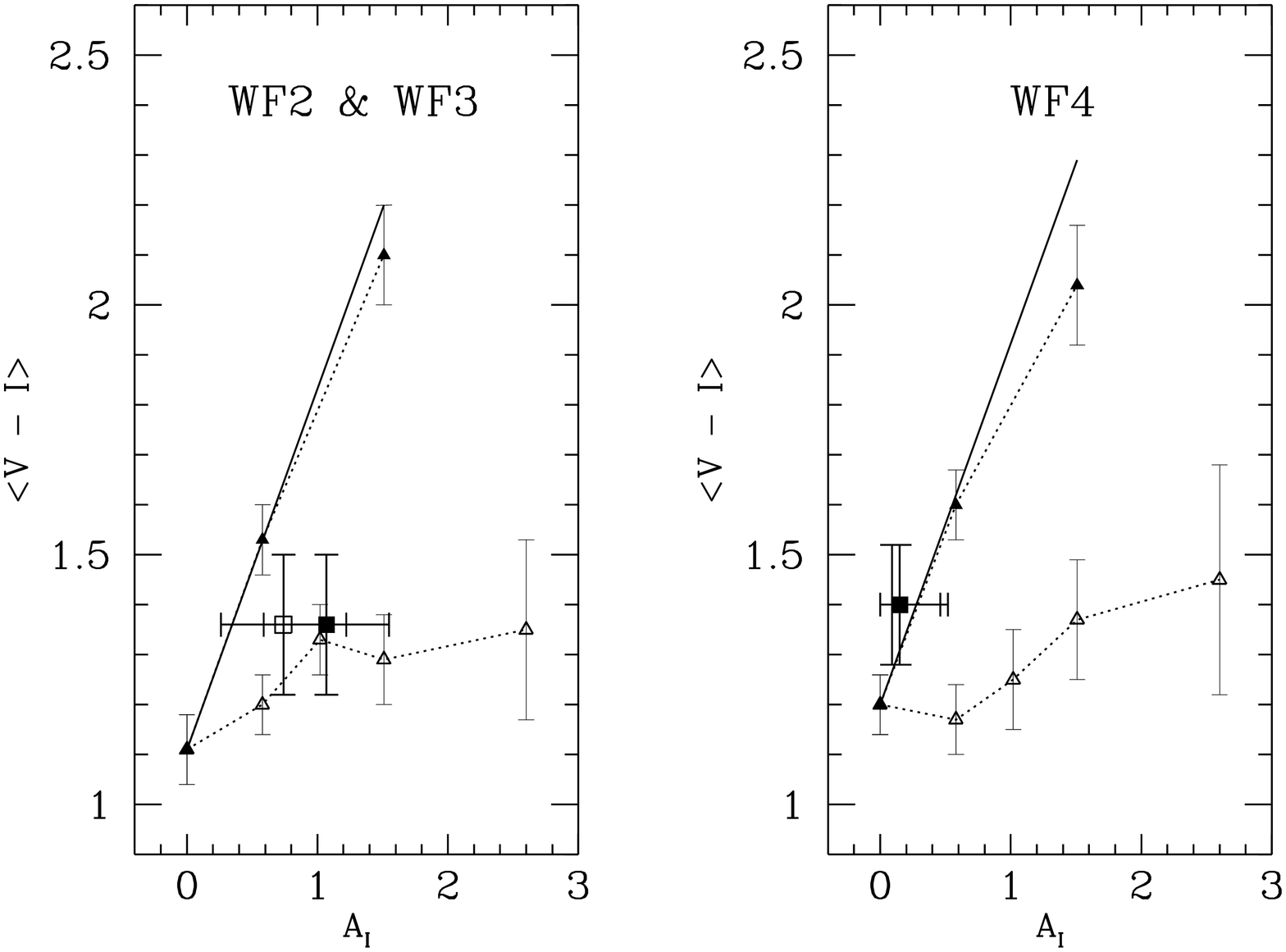,width=6.5in,angle=-90}
\caption{{\em Left: results for NGC 4536, WF2 and WF3
fields. Open triangles and dotted line:} 
dependence of the mean measured
color of the simulations on the
applied grey absorption.
{\em Filled triangles and dotted line:} mean measured
color of the simulations with extinction following
a Galactic reddening law.
{\em Solid line:} theoretical
(input) Galactic
reddening law. {\em Open square:} location
of the background galaxies behind
NGC 4536, derived from a grey absorbed model. {\em Filled square:} location
of real background galaxies behind NGC 4536, using a Galactic reddening law.
{\em Right: Results
for NGC 4536, WF4 field.}
Symbols as in left panel.}
\label{fred4536}
\end{figure}

\begin{figure}
\hskip 0.5in \psfig{figure=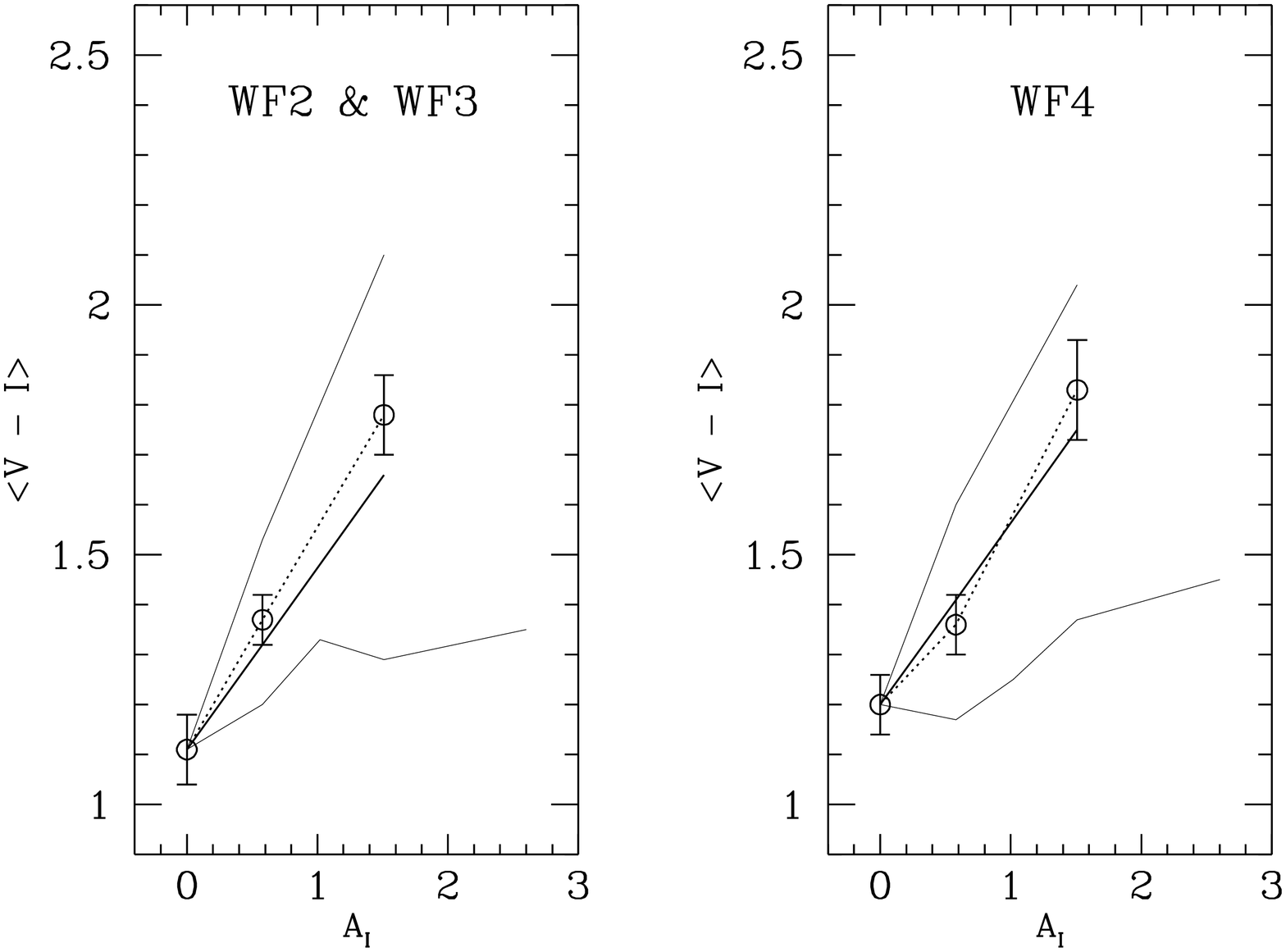,width=6.5in,angle=-90}
\caption{{\em Mixed simulations for NGC 4536. Left: 
WF2 and WF3.} 
{\em Open circles and thick-dotted line:}
mixed simulations. {\em Thick-solid line:} input of the 
mixed simulations. {\em Thin-solid lines:} results of 
Galactic reddening and grey simulations (Figure \ref
{fred4536}). {\em Right: WF4.} 
Symbols as in left panel.} 
\label{fmix4536}
\end{figure}

\begin{figure}
\plotone{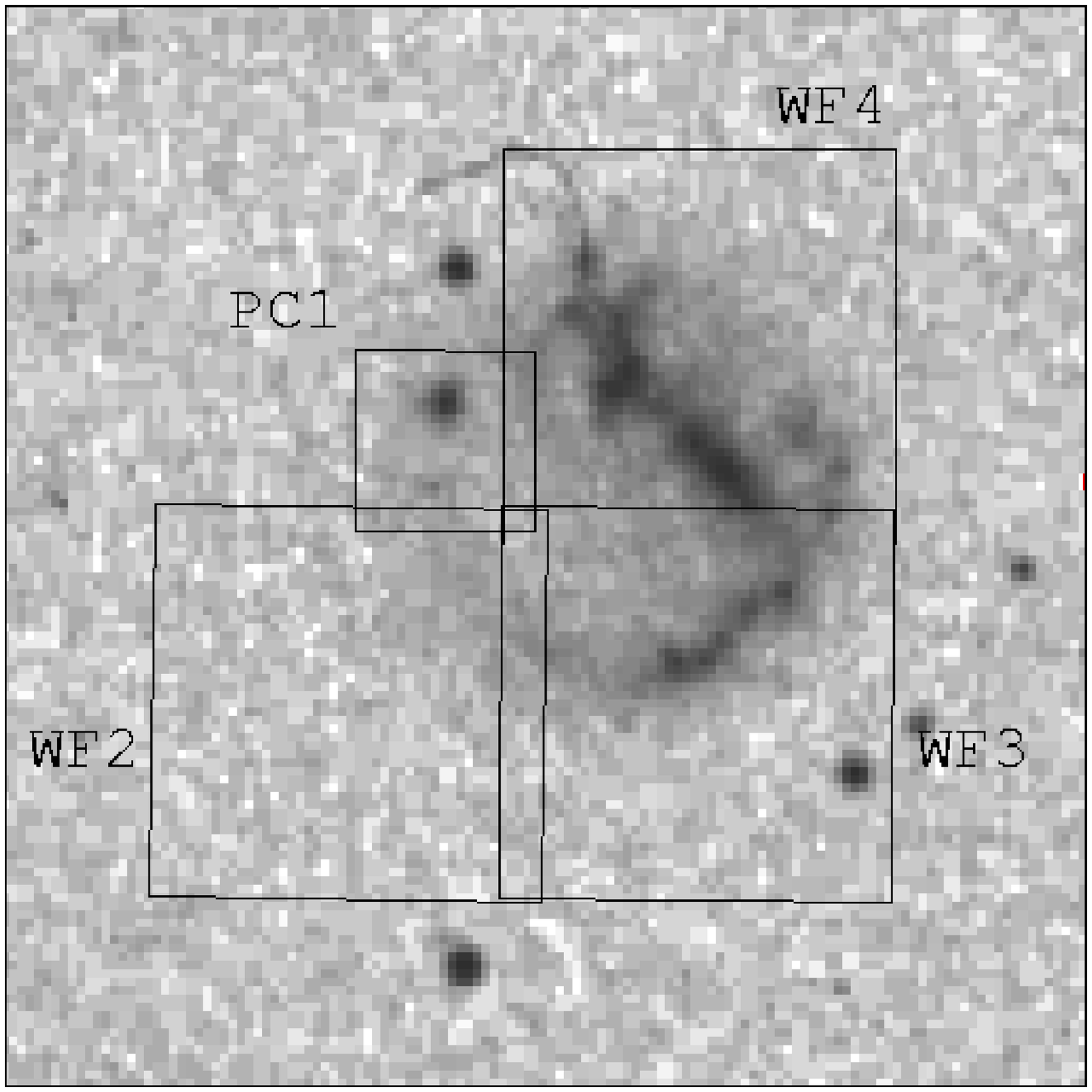}
\vspace*{0.2in}
\caption{STScI Digitized Sky Survey image of NGC 3664
displayed with a logarithmic scale.
The red (E) National Geographic Society--Palomar
Observatory Sky Survey plate was obtained on 1955 April 16,
with the 48-inch Oschin Schmidt Telescope.
The position of the
WFPC2 field is superimposed. 
North is up and East is to the left.
}
\label{fn3664}
\end{figure}

\begin{figure}
\plotfiddle{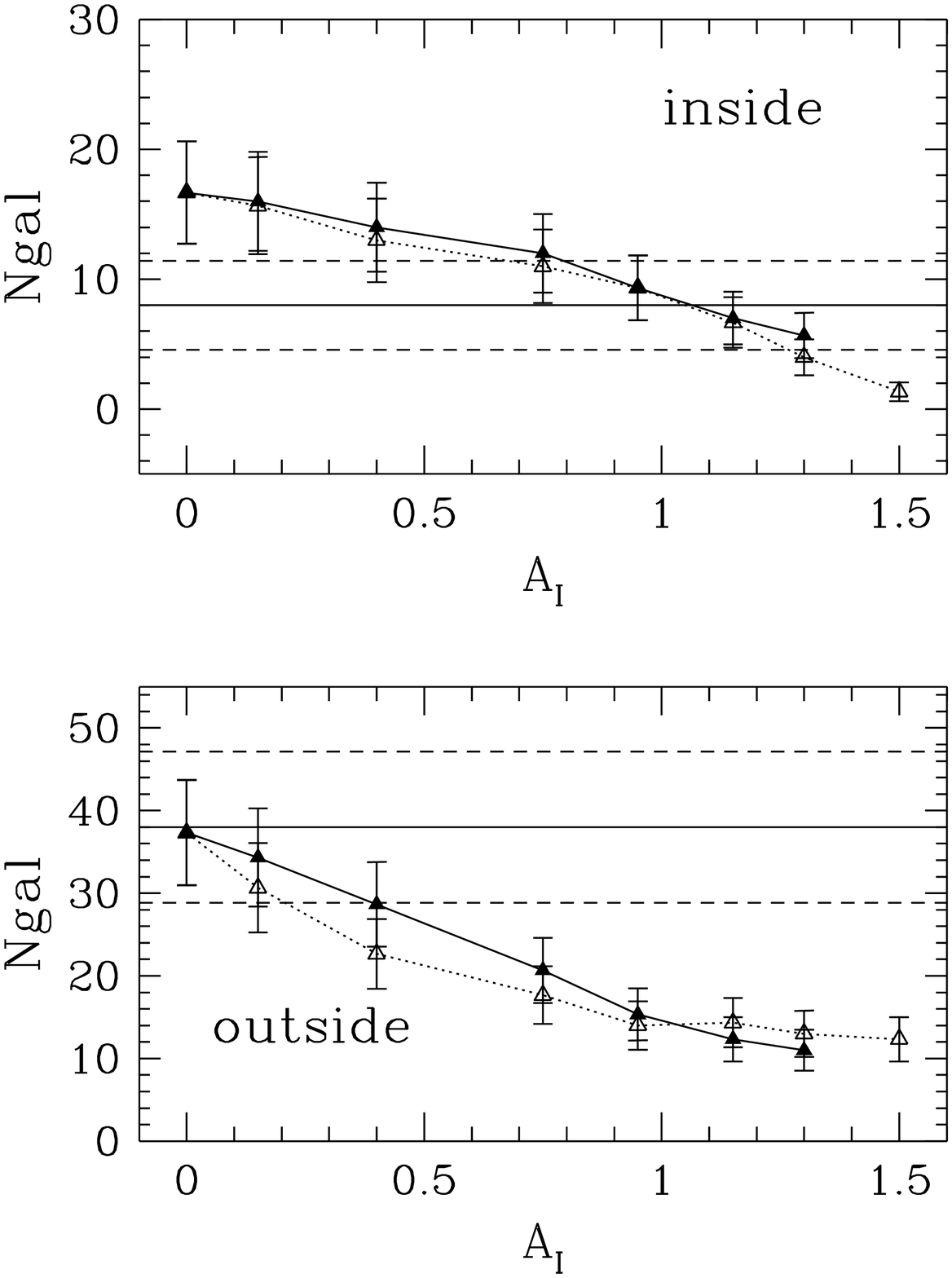}{6.3in}{0}{60}{60}{-210}{0}
\caption{{\em Top: results for the disk of NGC 3664.  Open 
triangles and dotted line:} average number of
recovered HDF galaxies 
vs. simulated grey extinction of $A_I$.
{\em Full triangles and solid line:} average number of recovered
HDF galaxies vs. extinction, when using a
Galactic reddening model for the simulations.
The error bars include Poisson uncertainty in the
number-counts and clustering error.
{\em Horizontal solid line:} number of real background galaxies
with its Poisson and clustering uncertainty
({\em horizontal short-dashed lines}).
{\em Bottom: results for the surrounding field.} Symbols as in top panel.}
\label{fext3664}
\end{figure}
 
\begin{figure}
\hskip 0.5in \psfig{figure=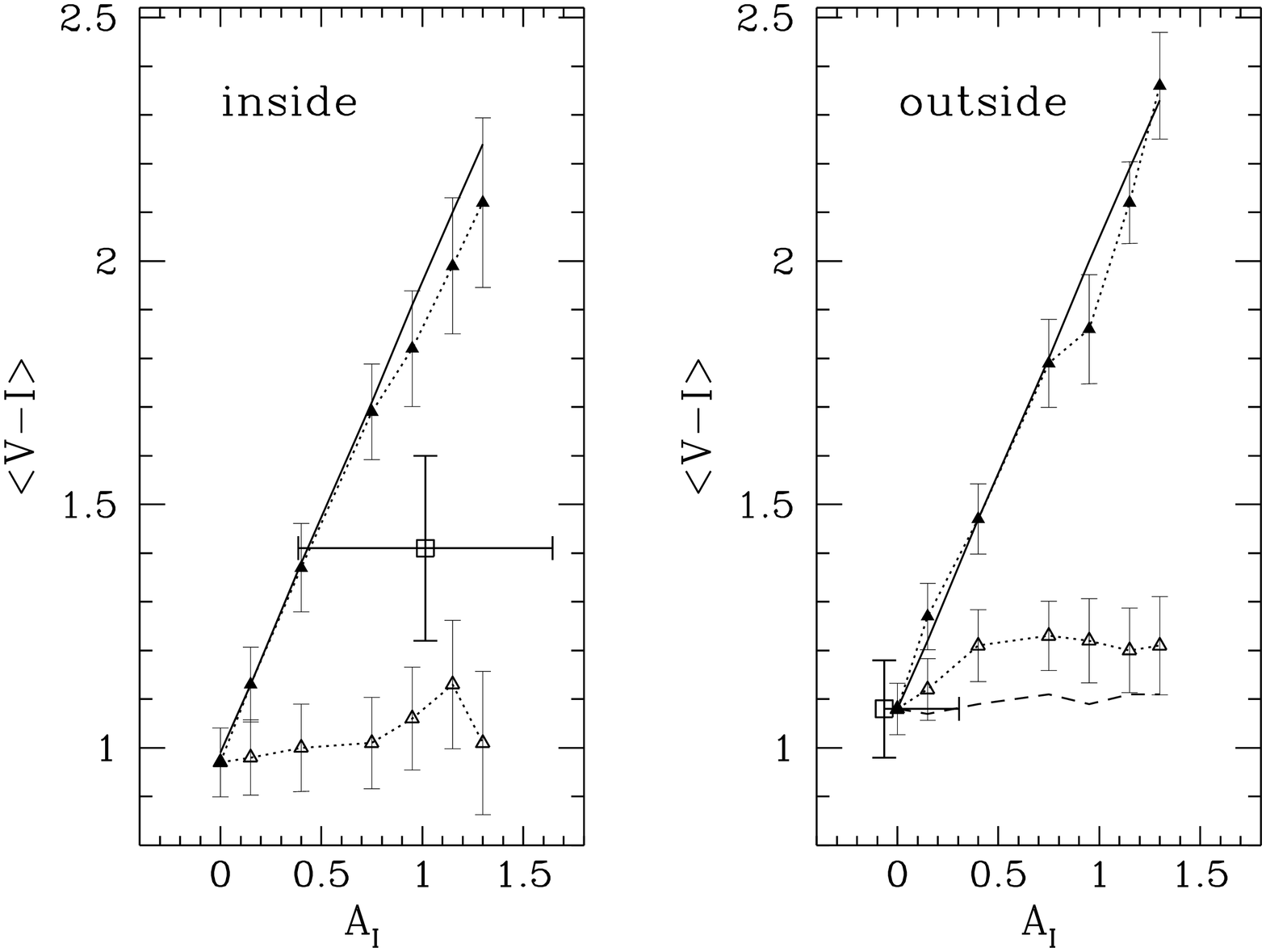,width=6.5in,angle=-90}
\caption{{\em Left: results for area inside the
disk of NGC 3664. Open triangles and
dotted line:} dependence of the mean measured
color on the
applied extinction for the simulations with grey absorption.
{\em Filled triangles and dotted line:} mean measured color 
of the simulations with Galactic reddening.
{\em Solid line:} theoretical
(input) Galactic
reddening law. {\em Open square:} location
of the background galaxies inside the disk area of
NGC 3664. 
{\em Right: results
for area outside of the disk of NGC 3664.}
{\em Long-dashed line:} measured colors of the
grey simulations, corrected for incompleteness.
Symbols as in left panel.}
\label{fred3664}
\end{figure}

\clearpage

\eject


\end{document}